\documentclass[conference]{IEEEtran}
\IEEEoverridecommandlockouts
\usepackage{cite}
\usepackage{amsmath,amssymb,amsfonts}
\usepackage{algorithmic}
\usepackage{graphicx}
\usepackage{textcomp}
\usepackage{xcolor}

\usepackage{subcaption}
\usepackage{enumitem}
\usepackage{ifthen}
\usepackage{soul}
\usepackage[frozencache=true,cachedir=.]{minted}

\usepackage{hyperref}

\usepackage{listings}
\usepackage{xspace}
\usepackage{booktabs}
\usepackage{multirow}
\usepackage{lscape} 
\usepackage{tcolorbox}

\def\BibTeX{{\rm B\kern-.05em{\sc i\kern-.025em b}\kern-.08em
    T\kern-.1667em\lower.7ex\hbox{E}\kern-.125emX}}

\newboolean{showcomments}
\setboolean{showcomments}{true}
\ifthenelse{\boolean{showcomments}}
{\newcommand{\nb}[2] {
\fcolorbox{black}{gray!20}{\bfseries\sffamily\scriptsize#1:}
{\sf\small$\blacktriangleright$\textit{#2}$\blacktriangleleft$}
}
}
{\newcommand{\nb}[2]{}
}

\newtheorem{termdefinition}{Definition}
\newcommand{\etal}{\textit{et al.}\xspace}
\newcommand{\tool}{\textsc{uncertainty-wizard}\xspace} %
\newcommand{\tfkeras}{\textit{tf.keras}\xspace} %

\begin{document}
\pagenumbering{arabic} 
\pagestyle{plain}

\title{Fail-Safe Execution of Deep Learning based Systems through Uncertainty Monitoring
\thanks{
This work was partially supported by the H2020 project PRECRIME,
funded under the ERC Advanced Grant 2017 Program (ERC Grant Agreement n. 787703).\newline
Accepted at ICST2021. 
© 2021 IEEE. Personal use of this material is permitted. Permission from IEEE must be
obtained for all other uses, in any current or future media, including
reprinting/republishing this material for advertising or promotional purposes, creating new
collective works, for resale or redistribution to servers or lists, or reuse of any copyrighted
component of this work in other works.
}
}

\author{\IEEEauthorblockN{Michael Weiss}
\IEEEauthorblockA{
\textit{Universit\`a della Svizzera italiana}\\
Lugano, Switzerland \\
michael.weiss@usi.ch}
\and
\IEEEauthorblockN{Paolo Tonella}
\IEEEauthorblockA{
\textit{Universit\`a della Svizzera italiana}\\
Lugano, Switzerland \\
paolo.tonella@usi.ch}
}


\maketitle

\begin{abstract}
Modern software systems rely on Deep Neural Networks (DNN) when processing complex, unstructured inputs, such as images, videos, natural language texts or audio signals.
Provided the intractably large size of such input spaces, the intrinsic limitations of learning algorithms  and the ambiguity about the expected predictions for some of the inputs, not only there is no guarantee that DNN's predictions are always correct, but rather developers must safely assume a low, though not negligible, error probability.
A fail-safe Deep Learning based System (DLS) is one equipped to handle DNN faults by means of a supervisor, capable of recognizing predictions that should not be trusted and that should activate a healing procedure bringing the DLS to a safe state.

In this paper, we propose an approach to use DNN uncertainty estimators to implement such supervisor. 
We first discuss advantages and disadvantages of existing approaches to measure uncertainty for DNNs 
and propose novel metrics for the empirical assessment of the  supervisor that rely on such approaches.
We then describe our publicly available tool \tool, which allows transparent estimation of uncertainty for regular \tfkeras DNNs.
Lastly, we discuss a large-scale  study conducted on four different subjects to empirically validate the approach,
reporting the lessons-learned as guidance for software engineers who intend to monitor uncertainty for fail-safe execution of DLS.
\end{abstract}


\section{Introduction}
\label{sec:introduction}

Deep neural networks (DNNs) are a powerful tool to identify patterns
in large amounts of data and to make predictions on new, previously unseen data.
Thanks to the increased hardware capabilities, DNNs can be run even on small, battery powered hardware
and can be trained in performance-optimized GPUs. Correspondingly, the use of DNNs has gained a lot of popularity in the last decade.
Moreover, the introduction of high level APIs such as \tfkeras (see \href{https://www.tensorflow.org/}{tensorflow.org})
allows even software engineers without previous experience in artificial intelligence to define, train and use custom DNNs.
DNNs are now used in many \emph{Deep Learning based Systems (DLS)}, like self driving cars, to interpret observed sensor measurements and control the car's actuators,
in medical systems, to support physicians to make their diagnosis, and in web services, for image processing and analysis.

Relying solely on the predictions made by a deep learning component might be dangerous, as there is always some \emph{uncertainty} about the correctness of the prediction. In fact, the \textit{contract} between the overall system and its DNN based components is necessarily a probabilistic one, and while the probability of an error can be low, it is never zero.

The uncertainty intrinsic with DNNs is either  caused by entropy in the input
or by inadequate training.
While the first type of uncertainty is inherent to a problem and cannot be avoided by definition,
the latter  cannot also be avoided for  practical reasons:
in most applications, the input space consists of a huge number of input contexts (e.g., the different weather or light conditions in which a car is driven), and it is impossible to collect data which perfectly represents all of them.

Faced with a problem for which a prediction is subject to high uncertainty, 
a human intelligence may consider to refuse to make a prediction and instead say `I do not know'.
DNNs on the other hand, will calculate a prediction for any given input, independently of the uncertainty of the prediction.
If such predictions are trusted by a DLS, the DLS may fail due to a wrong prediction, as is best illustrated by the following two examples: a self-driving car has recently crashed into an overturned truck. 
A likely explanation for such a crash is that overturned trucks are not sufficiently represented in the cars training data.\cite{Templeton2020Forbes}
Second, an online photo storage service classified an image of a black person as a a picture of a gorilla, 
leading to negative press, which deemed the service as racist. 
Again, such  error is likely caused by insufficient training data of the machine learning component which classified the image.\cite{Vincent2018GoogleFotos}
The fact that problems like these happen even in software from leading companies in the machine learning domain shows that preventing such errors is quite challenging.
Even more so, in the second example the solution put in place was a drastic workaround: 
the label \emph{Gorilla} was removed from the set of possible predictions for any input. 

We propose that DLS include a \emph{supervisor}, which monitors the DNN uncertainty for any given input at runtime,
such that the system can ignore predictions for high-uncertainty inputs and can run a safe fallback process instead, such as stopping the self-driving car at the side of the street or, 
in the second example, delegating the classification of the image to a human.

The machine learning community has investigated  \emph{uncertainty-aware} types of DNNs, which support the deployment of such a supervisor. 
This paper aims at closing the gap between uncertainty-aware DNNs and the deployment of an effective supervisor in a DLS. Specifically, it makes the following contributions:

\begin{description}[noitemsep]
\item [Metrics Comparison] Description and comparison of the most investigated uncertainty metrics for DNNs, with a discussion of their advantages and disadvantages.
\item [\tool] Python library which allows zero-knowledge, transparent implementation of uncertainty-aware DNNs.
\item [Evaluation Framework] We present existing and propose novel metrics to evaluate DLS which include a supervisor.
\item [Lessons learned] We discuss key findings from our empirical evaluation of various uncertainty metrics applied to four different case studies.
\end{description}

\section{Background}
\label{sec:background}

In this section, we discuss the different root causes of DNN faults which can be understood as \emph{types of uncertainty}
and define the task of DNN fault prediction as a problem of uncertainty quantification.

\subsection{Sources of Uncertainty}

We distinguish between two types of uncertainty; uncertainty caused by a sub-optimal DNN model and uncertainty caused by randomness in the prediction target. 
A detailed discussion of these types is provided by Kendall \etal \cite{Kendall2017a}.

\begin{termdefinition}[Epistemic Uncertainty]
    Epistemic uncertainty is caused by the sub-optimal training or configuration of the \emph{model}.
\end{termdefinition}

Epistemic uncertainty is sometimes also referred to as \emph{model uncertainty}.
There are many possible reasons for epistemic uncertainty, such as 
insufficient training data, which does not  represent the entire possible input space,
sub-optimal training hyper-parameters and inadequate DNN architecture. 
In theory, epistemic uncertainty could be avoided, 
provided good enough training data and optimal model configuration.
However, finding such optimal training configurations and data is impossible in most real world applications,
as real input spaces, as well as the space of the possible hyper-parameters and architectural choices, are typically too large.

The second type of uncertainty, which not even an optimal training set and model configuration can avoid, is called \emph{aleatoric uncertainty}:

\begin{termdefinition}[Aleatoric Uncertainty]
    Aleatoric uncertainty is the uncertainty present in the true (unknown) distribution we are making predictions about.
\end{termdefinition}

Thus, aleatoric uncertainty can be seen as randomness, ambiguity or entropy in the prediction target.
When predicting a random event, even an optimal model will make wrong predictions. 
As aleatoric uncertainty is independent of the model, but instead depends on the predicted data, it is also referred to as \emph{data uncertainty} or \emph{irreducible uncertainty}.
Aleatoric Uncertainty can  be further distinguished between \emph{homoscedastic uncertainty}, 
where the uncertainty applies to all data, and \emph{heteorscedastic uncertainty}, 
where the uncertainty is more prevalent amongst some subsets of the data.


\def \unctypesubfigurewidth {0.18\linewidth}
\def \unctypeimgwidth {.8\linewidth}

\begin{figure}
\centering
\begin{subfigure}{\unctypesubfigurewidth}
  \centering
  \includegraphics[width=\unctypeimgwidth]{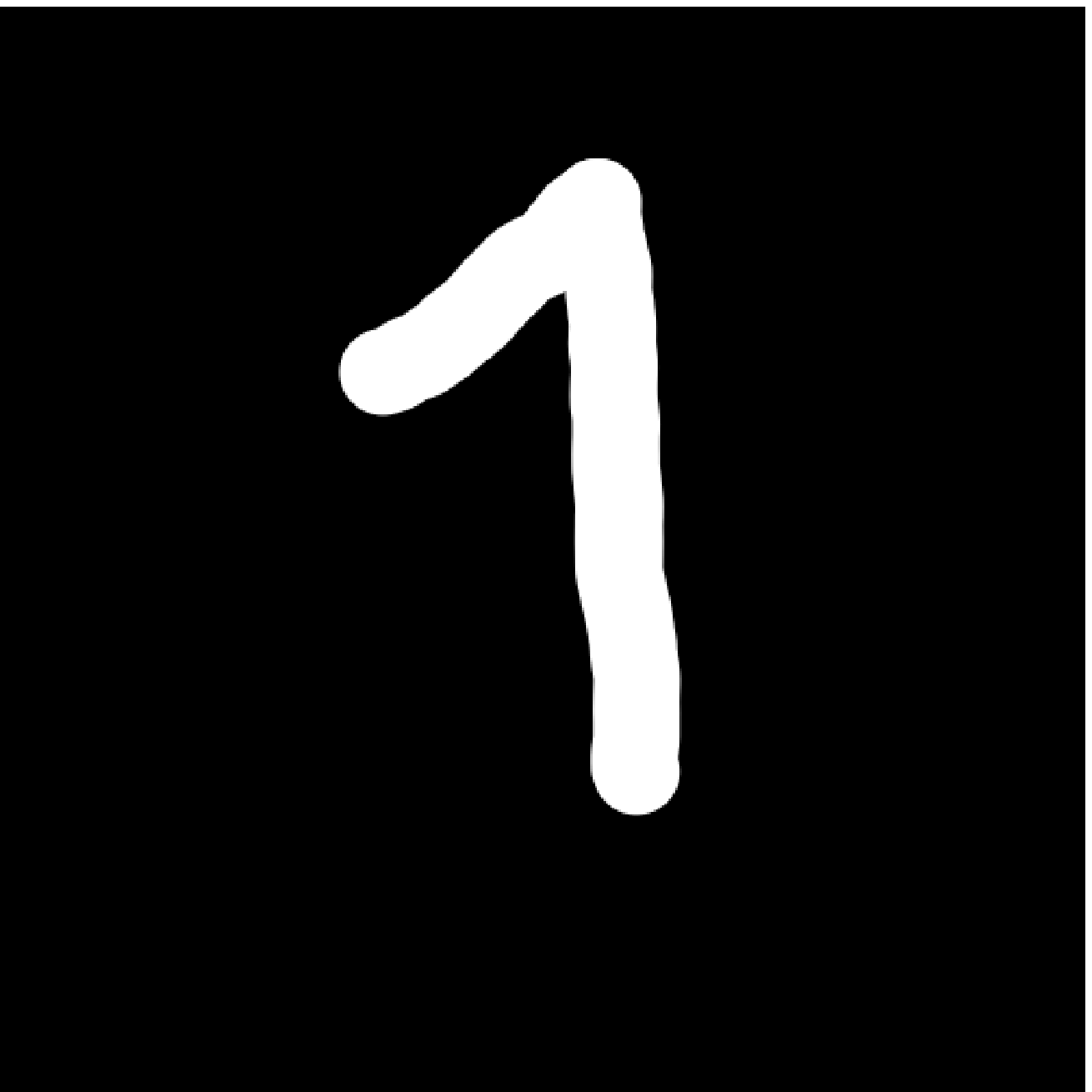}
  \caption{}
  \label{fig:u_type_nominal_1}
\end{subfigure}%
\begin{subfigure}{\unctypesubfigurewidth}
  \centering
  \includegraphics[width=\unctypeimgwidth]{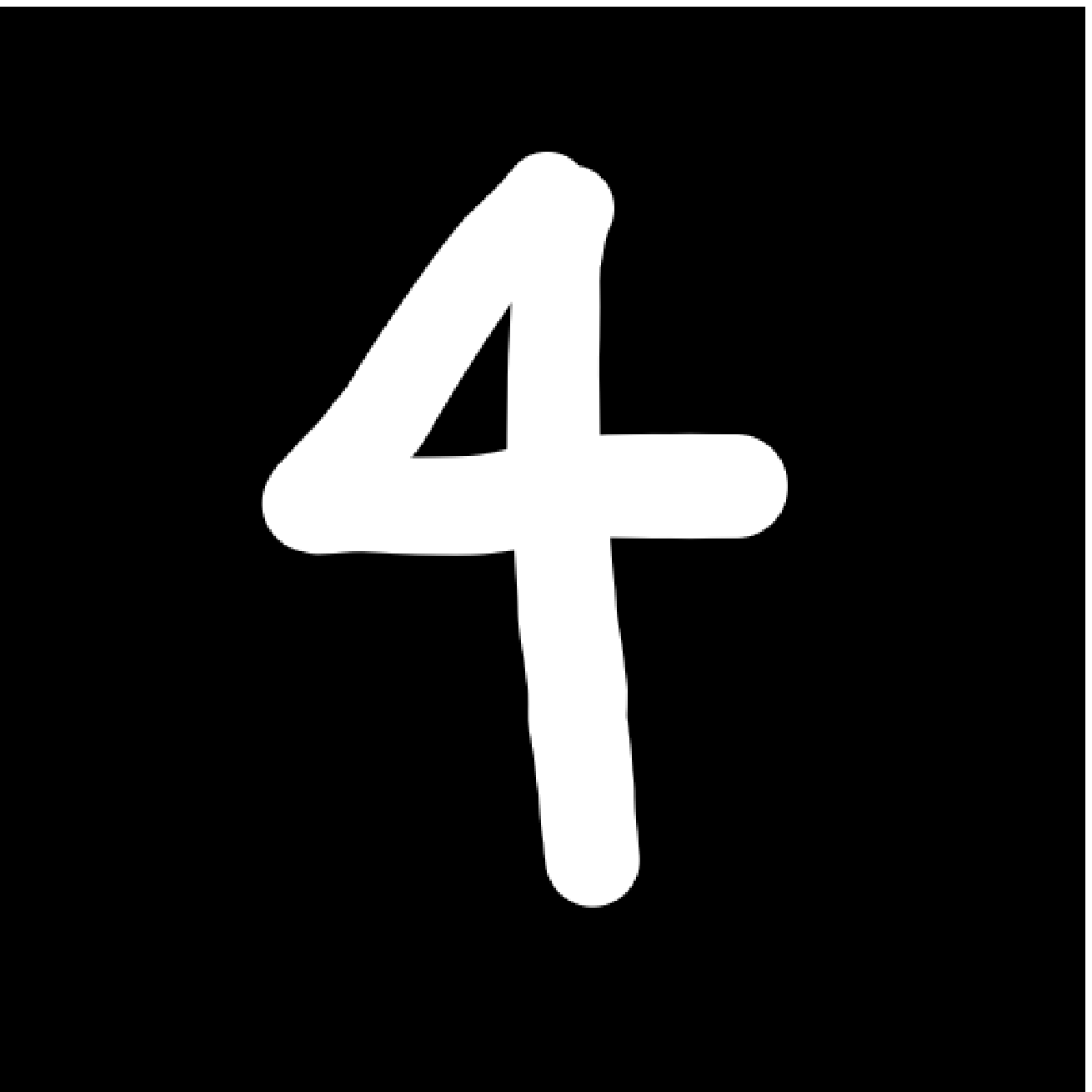}
  \caption{}
  \label{fig:u_type_nominal_4}
\end{subfigure}
\begin{subfigure}{\unctypesubfigurewidth}
  \centering
  \includegraphics[width=\unctypeimgwidth]{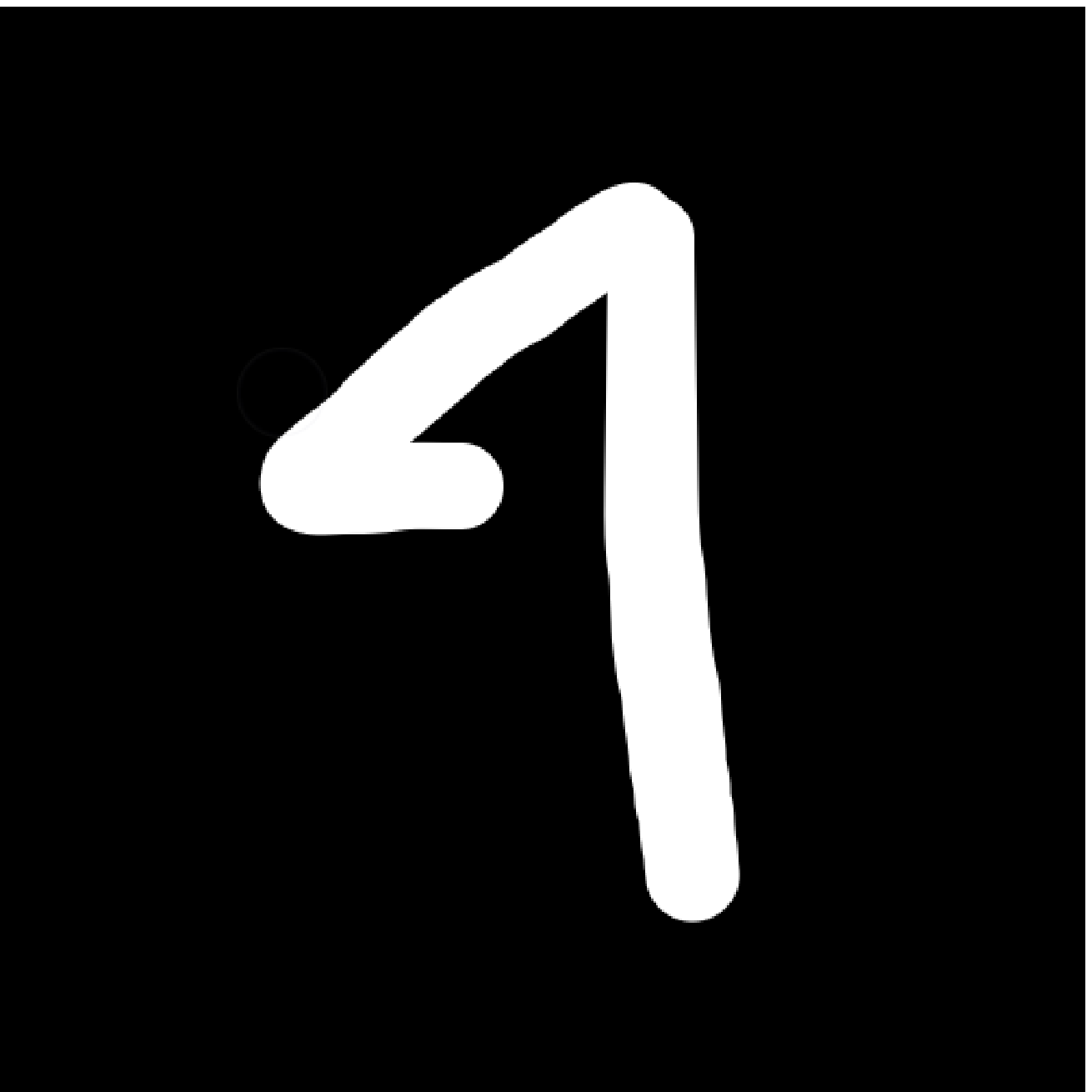}
  \caption{}
  \label{fig:u_type_aleatoric}
\end{subfigure}%
\begin{subfigure}{\unctypesubfigurewidth}
  \centering
  \includegraphics[width=\unctypeimgwidth]{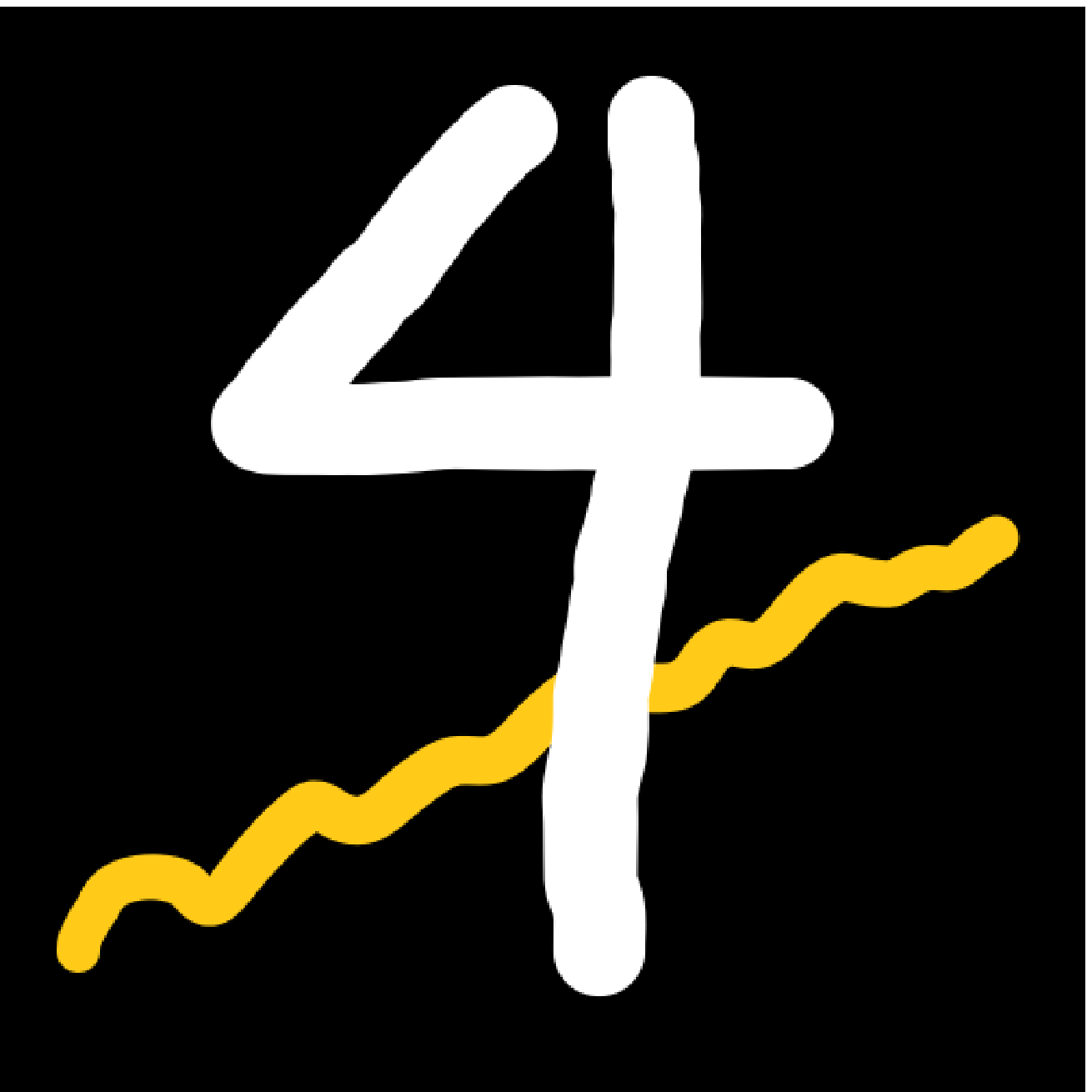}
  \caption{}
  \label{fig:u_type_epistemic_corrupted}
\end{subfigure}
\begin{subfigure}{\unctypesubfigurewidth}
  \centering
  \includegraphics[width=\unctypeimgwidth]{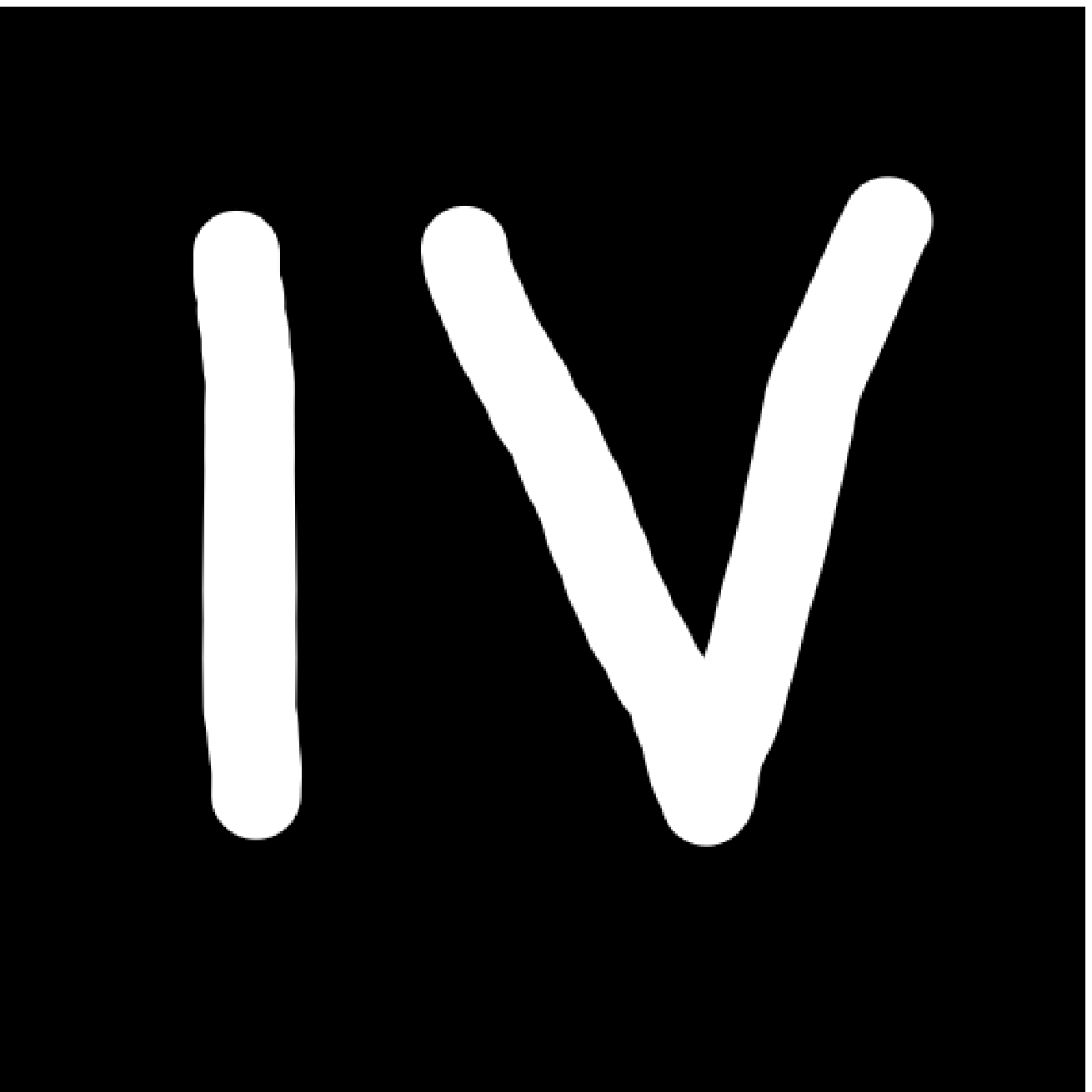}
  \caption{}
  \label{fig:u_type_epistemic_roman}
\end{subfigure}

\caption{
Examples of uncertainties in digits classification: 
(\subref{fig:u_type_nominal_1}) and (\subref{fig:u_type_nominal_4}) cause no uncertainty,
(\subref{fig:u_type_aleatoric}) causes aleatoric uncertainty, 
(\subref{fig:u_type_epistemic_corrupted}) and (\subref{fig:u_type_epistemic_roman}) cause epistemic uncertainty.
}
\label{fig:uncertainty_types}
\end{figure}
Figure \ref{fig:uncertainty_types} provides a visual example of the difference between aleatoric and epistemic uncertainty.
The input to a classifier DNN is the image of a handwritten digit to be recognized.
Figures \ref{fig:u_type_nominal_1} and \ref{fig:u_type_nominal_4} illustrate regular inputs with low uncertainty.
Figure \ref{fig:u_type_aleatoric} is an example of a figure with high heteroscedastic aleatoric uncertainty: 
Clearly, the image shows either a 1 or a 4, but it is impossible to say with certainty which one the writer intended.
Figure \ref{fig:u_type_epistemic_corrupted} illustrates a common reason for epistemic uncertainty:
A small perturbation of the image background, if not present in the training data, may lead the model to 
be incapable of predicting the correct label. 
Similarly, \ref{fig:u_type_epistemic_roman} shows an unexpected input leading to epistemic uncertainty.
While the true label is unambiguously 4, a model which was not trained on roman number representations will not be capable 
to make a correct prediction.

\subsection{Uncertainty Quantification}
Ideally, instead of predicting a single value as output, a DNN should calculate a probability density function for regression problems or a likelihood for every outcome in a classification problem.
As such, every outcome would have its uncertainty quantified (e.g., by the variance of the output probability distribution).
We will discuss models capable of calculating such outputs in Section \ref{sec:approaches}.
However, for the scope of this paper, we consider a less general formulation of uncertainty quantification,
which is sufficient for network supervision~\cite{Riccio2020}:

\begin{termdefinition}[Uncertainty Quantification]
    Uncertainty Quantification (UQ) is the task of calculating a scalar metric
    strictly monotonically increasing in the likelihood (for classification tasks) or severity (for regression tasks)
    of a deviation between the DNN prediction and the ground truth, given a particular input and the DNN used to do the prediction.
\end{termdefinition}

We are thus limiting our interest to the correctness of the chosen prediction,
as opposed to the distribution of all possible predictions in the output space.
The supervisor will reject inputs for which the uncertainty is above a certain threshold.

Consistently, we define \emph{confidence} as the opposite of uncertainty, 
s.t. a confidence metric is supposed to strictly monotonically decrease in the likelihood or severity of a prediction error.

\section{Uncertainty-Aware Neural Networks}
\label{sec:approaches}
\def \spacebetweenapproachrows {0.1cm}

\begin{table*}[t]
\centering
\begin{tabular}{@{}lllllllll@{}}
\toprule
\textbf{\begin{tabular}[c]{@{}l@{}}Uncertainty-\\ aware DNN\end{tabular}}            & \textbf{\begin{tabular}[c]{@{}l@{}}Classifi-\\ cation\end{tabular}} & \textbf{\begin{tabular}[c]{@{}l@{}}Regre-\\ ssion\end{tabular}}& \textbf{BNN} & \textbf{\begin{tabular}[c]{@{}l@{}}Custom\\ Layers\end{tabular}} & \textbf{\begin{tabular}[c]{@{}l@{}}Training\\ Effort\end{tabular}} & \textbf{\begin{tabular}[c]{@{}l@{}}Prediction\\ Effort\end{tabular}} & \textbf{\begin{tabular}[c]{@{}l@{}}Main \\ Advantage\end{tabular}}                                     & \textbf{\begin{tabular}[c]{@{}l@{}}Main \\ Disadvantage\end{tabular}}                                              \\ \midrule

\vspace{\spacebetweenapproachrows}
\textbf{\begin{tabular}[c]{@{}l@{}}Variants of \\ pure BNN\end{tabular}}                                                                                                                                   & Yes                                                             & Yes                                                                          & Yes          & \begin{tabular}[c]{@{}l@{}}Weight \\ distributions\end{tabular}  & \begin{tabular}[c]{@{}l@{}}High or\\ intractable*\end{tabular}      & \begin{tabular}[c]{@{}l@{}}High or\\ intractable*\end{tabular}        & \begin{tabular}[c]{@{}l@{}}Theoretically \\ well founded\end{tabular}                                         & \begin{tabular}[c]{@{}l@{}}Custom architecture, \\ computationally hard*\end{tabular}                                          \\
\vspace{\spacebetweenapproachrows}
\textbf{MC Dropout}                                                                                                                                & Yes                                                             & Yes                                                                          & Yes          & Dropout                                                          & Minimal                                                                & High                                                                 & \begin{tabular}[c]{@{}l@{}}Fastest BNN\\ approximation\end{tabular}                                    & \begin{tabular}[c]{@{}l@{}}Sampling for \\ predictions is costly\end{tabular}                                      \\
\vspace{\spacebetweenapproachrows}
\textbf{\begin{tabular}[c]{@{}l@{}}Deep \\ Ensembles\end{tabular}}                                                                                 & Yes                                                             & Yes                                                                          & No*           & No                                                               & High                                                               & High                                                                 & \begin{tabular}[c]{@{}l@{}}Good practical results,\\ no major architecture\\ requirements\end{tabular} & \begin{tabular}[c]{@{}l@{}}Computationally and \\ storage intensive\end{tabular}                                              \\
\vspace{\spacebetweenapproachrows}
\textbf{\begin{tabular}[c]{@{}l@{}}PPNN-Softmax\\ based\end{tabular}} & Yes                                                                 & No                                                                                                                                 & No           & Softmax                                                          & Miminal                                                            & Minimal                                                              & Fast and Simple                                                                                        & \begin{tabular}[c]{@{}l@{}}Misleading and\\ no regression\end{tabular}                                                \\
 \bottomrule
\end{tabular}
\caption{Overview of popular uncertainty-aware DNN techniques (* = approach dependent) }
\end{table*}
\label{tab:supervision_approaches}

While regular DNNs, also called \emph{Point-Prediction DNNs (PPNN)}, predict a single scalar value for every output variable, from the perspective of uncertainty quantification it would be preferable if the DNN calculated a distribution over possible output values, which would allow the extraction of the network's confidence in the prediction. 
In this section we discuss the various DNN architectures which allow the calculation of such distributions:
first, we  provide an overview on \emph{Bayesian Neural Networks (BNN)}, which are the standard approach in the machine learning literature about uncertainty quantification.\footnote{
For a more precise discussion of BNN, we recommend the comprehensive and usage-oriented article by Jospin \etal ~\cite{Jospin2020}.
}
Second, we discuss other approaches, which typically have some theoretical disadvantages when compared to BNNs, but showed good results in practice.\cite{Ovadia2019}

\subsection{Pure Bayesian Neural Networks}
BNNs are neural networks in which weights are probabilistic, instead of scalar as in PPNN, and are represented as probability density functions.
To train a BNN, first, a prior distribution $p(\theta)$ over weights $\theta$ has to be defined. 
Then, given some data $D$, the posterior distribution $p(\theta|D)$, i.e., the trained BNN is inferred using Bayes rule: 
\begin{equation}
p(\theta|D) = \frac{p(D|\theta) p(\theta)}{p(D)}  
= \frac{p(D|\theta) p(\theta)}{\int p(D|\theta) p(\theta) d\theta}
\end{equation}

Since weights are probabilistic, any  output of the BNN is also probabilistic and thus allows a statistically well-founded uncertainty quantification.
Besides that, these BNNs in their pure form  have other advantages over PPNN, among which: 
they are robust to overfitting~\cite{Neal1992} and allow to distinguish between epistemic and aleatoric uncertainty~\cite{Jospin2020}.

BNNs in the pure form presented above quickly become intractable due to the large number of integrations required.
There has been a long search for mechanisms which make BNN easier to compute, with  papers dating back to the early 1990s~\cite{MacKay1992, Neal1992}. 
To do so, some approaches make additional assumptions, e.g., a normal distribution of the weights, which may cause the BNN to lose some of the advantages listed above.
Still, despite many advances in the field, even recent approaches are  considered not to scale  sufficiently well~\cite{Ilg2018} and are not yet widely used in practice~\cite{Jospin2020}.
Other, more scalable approaches have been shown to  outperform Bayesian approaches~\cite{Ovadia2019} on practical benchmarks.

\subsection{MC-Dropout based Bayesian Neural Networks}
In their very influential paper\footnote{More than 2300 citations in the four years since its publication, according to \href{https://scholar.google.com/}{Google Scholar}.}, Gal \etal~\cite{Gal2016} proposed to approximate BNNs though regular PPNNs.
Many PPNNs use \emph{dropout layers} for regularization. During training, in a dropout layer each neuron activation can be set to 0 with some probability $p$. This helps to avoid overfitting and can lead to better model performance in general~\cite{Srivastava2014}.
In their paper, Gal \etal have shown that a PPNN trained with dropout enabled can be interpreted as a BNN,
due to the variation induced by the randomized dropout layers.
While traditionally the dropout functionality is disabled at prediction time, a dropout-based BNN keeps the random dropping of neuron activations
enabled even during predictions and calculates an output distribution by Monte-Carlo sampling of the results in multiple randomized  predictions.
This approach is thus often referred to as \emph{MC-Dropout}.

Despite its high popularity, MC-Dropout  is not without criticisms: 
Osband \cite{Osband2016} claimed the inability of MC-Dropout to capture every type of uncertainty and
several papers have shown that it can be outperformed by other approaches~\cite{Lakshminarayanan2017, Ovadia2019}.
Also, despite the fact that \textit{"the forward passes can be done concurrently, resulting in constant running time"} \cite{Gal2016},
concurrent processing may not be possible in resource-constrained environments, making MC-Dropout clearly slower than a single prediction of the corresponding PPNN.

\subsection{Deep Ensemble Neural Networks}
Lakshminarayanan \etal proposed a fundamentally different approach to quantify uncertainty called \emph{Deep Ensembles}, or \emph{Ensemble} for short, i.e., a collection of multiple \emph{atomic models} for the same problem,
differing only in their  initialization and trained independently. 
For every input, a prediction would be made on every atomic model. 
The predictive distribution could then be inferred from these samples.
While deep ensembles are not inherently BNNs, it is possible to interpret them as BNN after applying some minor changes to the parameter regularization~\cite{Pearce2020}.
Nonetheless, even in their plain form, they have been shown to outperform MC-Dropout on the task of uncertainty quantification~\cite{Lakshminarayanan2017, Ovadia2019}.

Deep Ensembles are, compared to a single PPNN, slow to train if executed sequentially and memory intensive if used concurrently.
This may prevent the use of ensembles in some constrained environments.
A variety of improvements and modifications have been proposed for Deep Ensembles (Ilg \etal \cite{Ilg2018} provide a good overview of them).

\subsection{Point Predictor Classifiers}
DNNs used for classification typically have a \emph{softmax} output layer, including one output neuron per class. 
The output for each class is between $0$ and $1$, with the sum of all outputs being exactly $1$.
These outputs are often interpreted as probability distributions over the classes
and are used for network supervision by means of the following quantifiers:

\begin{termdefinition}[Max-Softmax (SM)] The highest softmax score is used as confidence quantification 
   (also referred to as \emph{Vanilla}~\cite{Ovadia2019} or \emph{Softmax Prediction Probability}~\cite{Berend2020} quantification).
    
\end{termdefinition}

\begin{termdefinition}[Prediction Confidence Score (PCS)~\cite{Zhang2020}]
    The difference between the two highest softmax outputs is used as confidence quantification.
\end{termdefinition}

\begin{termdefinition}[Softmax-Entropy (SME)]
    The entropy over the outputs of the softmax layer is used as  confidence quantification.
\end{termdefinition}

These quantifiers are often criticized as a poor approach to uncertainty quantification:
As opposed to BNNs and BNN approximations, PPNN based quantifiers are not theoretically well founded,
and can be shown to severely overestimate a network's confidence~\cite{Gal2016}.
As a further disadvantage, such a PPNN based approach can not be directly applied to regression problems.

\subsection{Inferring Prediction and Uncertainty from Samples}
In MC-Dropout and Deep Ensembles, samples need to be aggregated into a point prediction and uncertainty quantification,
and the literature provides a variety of quantifiers able to do so.
In this Section, we first discuss the quantifiers applicable to regression problems, and then the ones for classification problems.

\paragraph{Regression Problems}

In their proposition of MC-Dropout, Gal \etal~\cite{Gal2016} propose to use the average of the observed samples as prediction,
and their predictive variance as uncertainty:
\begin{termdefinition}[Predictive Variance]
    The \emph{predictive variance} is the sample variance over the observed samples plus the inverse model precision.
\end{termdefinition}
The inverse model precision is constant for a given model. 
Thus, for the purpose of network supervision, where strictly monotonic transformations of uncertainty quantification scores
do not change the supervisor performance, using  predictive variance is equivalent to using the sample variance or the sample standard deviation.
An alternative approach was proposed by Lakshminarayanan \etal\cite{Lakshminarayanan2017}.
For their deep ensembles, they propose that the atomic models are adapted s.t. they have a second output variable for every regression output which predicts the variance~\cite{Nix1994}. 
The uncertainty of the ensemble can then be quantified by averaging these variances. 

\paragraph{Classification Problems}
The following quantifiers are proposed to derive an overall prediction and uncertainty:

\begin{termdefinition}[Mean-Softmax (MS)]
    The overall prediction is the class with the highest sum of softmax scores over all samples
    and the corresponding confidence is the average softmax score of this class over all samples.
\end{termdefinition}

MS has been proposed and is often used with Deep Ensembles. It is thus also called \emph{ensembling}\cite{Jospin2020}.
Three other quantifiers have been proposed to be used with MC-Dropout\cite{Gal2016a, Gal2016, Michelmore2018}:
\emph{Variation Ratio (VR)}, which is defined as the percentage of samples for which the overall chosen class is \textbf{not} the class with the highest softmax output; \emph{Predictive Entropy (PE)}, which measures the average amount of information in the predicted distributions;
and the \emph{Mutual Information (MI)} between a prediction and the models posterior (see 
Gal, 2016~\cite{Gal2016}, Section 3.3.1 for a precise description and for examples of these three uncertainty quantifiers).

\section{\tool}
\label{sec:wizard}

\begin{listing}[t]
    \begin{minted}[fontsize=\footnotesize]{python}
    model = wizard.models.StochasticSequential()
    # Use as a regular tf.keras model
    model.add(tf.keras.layers.Dense(100)
    model.add(tf.keras.layers.Dropout(0.2)
    model.add(tf.keras.layers.Softmax(10)
    model.compile( ... )
    model.fit( ... )
    # Use as standard (non-wizard) model
    regular_nn_output = model.predict(x_test)
    # Predict as Point-Predictor w/ confidence
    pred_pp, pcs = model.predict_quantified(x_test,
        quantifier='pcs')
    # Predict as Bayesian Model w/ uncertainty
    pred_b, var_r = model.predict_quantified(x_test,
        num_samples=100, quantifier='var_ratio')
    \end{minted}
    \caption{Keras-Syntax Stochastic Model}
    \label{lst:stochastic_sequential_snippet}
\end{listing}

\begin{listing}[t]
    \begin{minted}[fontsize=\footnotesize]{python}
    model = ... # load or create plain keras model
    model = wizard.models.stochastic_from_keras(model)
    # model can now be used as point-predictor
    # and as bayesian model, i.e.,
    model.predict(...)
    model.predict_quantified(...) 
    \end{minted}
    \caption{Convert Pre-Trained Model}
    \label{lst:from_keras_snippet}
\end{listing}

With  148'742 stars and 82'784 forks on github.com, Tensorflow is presumably the most popular deep learning framework.\footnote{Its main alternative, \emph{PyTorch} has 42'621 stars and 11'102 forks.}~\cite{TfPytorchCompare}
In its recent versions, a large part of its API, \tfkeras, is based on the popular Keras API, a simple yet powerful high-level API to develop, train and deploy DNNs.
The simplicity of the \tfkeras API allows researchers and practitioners outside of the machine learning community to get started with deep learning easily.
Unfortunately, such simple API does not expose equally simple methods to quantify uncertainty.
Thus, we release \tool, an extension of \tfkeras which allows developers to easily create Stochastic and Ensemble DNNs
and apply all uncertainty quantifiers described in Section \ref{sec:approaches}.
The core features of \tool are:
\begin{description}[noitemsep]
\item [Sequential API] Sequential models are the most straightforward way to use \tfkeras
and are thus very popular. 
However,  the sequential API does not allow dropout at prediction time. Hence, it also does  not allow the implementation of MC-Dropout.
\tool closes this gap by supporting the creation of stochastic models using  the sequential, as well as  the functional, API
in plain \tfkeras syntax. An example of this is given in \autoref{lst:stochastic_sequential_snippet}.
\item [Dynamic Randomness] In \tfkeras, the dropout behavior at prediction time is unchangeable for a given model: 
Either it is disabled (as required in point predictors) or enabled (as required in stochastic models). 
Thus, despite relying on the same architecture and weights, a \tfkeras model cannot be used both as stochastic model and as point predictor.
Converting them is nontrivial and includes the creation of a new model, which is memory and performance intensive.
\tool's sequential models dynamically enable and disable stochastic behavior based on whether the passed quantifier
expects a point prediction or sampled predictions. 
This is shown in \autoref{lst:stochastic_sequential_snippet} as well.
\item [Conversion from Keras] Often, the user of a DNN is not the same person or group that trained the model.
To allow users to use such a model for MC-Dropout nonetheless,
\tool supports the import of any \tfkeras model (which has at least one dropout layer) as a stochastic model.
\item [Parallelized Ensembles] \tfkeras API does not expose simple functionality for parallel training of multiple models. 
Especially with smaller models, which do not require the full use of the existing hardware to be loaded and executed,
sequential training of an ensembles atomic models has a large negative impact on training and prediction time.
Additionally, it can also lead to pollution of the global tensorflow runtime
due to memory leaks and eager processing.\footnote{See e.g. tensorflow issues \href{https://github.com/tensorflow/tensorflow/issues/33030}{33030}
and \href{https://github.com/tensorflow/tensorflow/issues/37505}{37505}.}
\tool treats ensembles lazily: every atomic model is stored on the file system,
and lazily loaded into its own tensorflow runtime during execution.
This allows faster, parallelized execution without runtime pollution.
\item [Dependency-Light \textit{pip install}] \tool is platform independent, 
importable
through \textit{pip install uncertainty-wizard}
and has only one dependency: Tensorflow version 2.3.0 or later. 
\end{description}

Due to space constraints, the description of \tool at this place is brief. 
We provide a more extensive discussion in a technical tool paper \cite{Weiss2020Wizard}.
\tool and a comprehensive user guide can be found online: 
    \textbf{\href{https://github.com/testingautomated-usi/uncertainty-wizard}{github.com/testingautomated-usi/uncertainty-wizard}}

\section{Supervised Neural Network Assessment}
\label{sec:assessment}

Network supervision can be viewed as a binary classification task: 
\emph{malicious samples}, i.e., inputs which lead to a misclassification (for classification problems) or to severe imprecision (in regression problems) are positive samples that have to be rejected.
Other samples, also called \emph{benign samples}, are negative samples in the binary classification task.
An uncertainty based supervisor accepts an input $i$ as a benign sample if its uncertainty $u(i)$ is lower than some threshold $t$.
The choice of $t$ is a crucial setting, as a high $t$ will fail to reject many malicious samples (false negatives) and a low $t$ will cause too many false alerts (false positives).

Differently from standard binary classification tasks, the choice of $t$ for network supervision cannot rely on the optimization of an aggregate metric that accounts for both false positives and false negatives, such as the F1-metric, because negative samples are either completely unknown at training time, or, in case some negative samples are known, they cannot be assumed to be representative of all unknown execution conditions that will give raise to uncertainty at runtime.
Hence, the choice of $t$ is solely based on the false positives observed in the validation set.
In practice, given the uncertainties measured on the validation set, $t$ shall ensure that at runtime under similar conditions only an acceptable \emph{false positive rate} $\epsilon$ is expected to occur~\cite{Stocco2020}.

Existing metrics allow the individual assessment of the supervisor's performance separately from the assessment of the model performance\cite{Henriksson2019}. However, such measurements do not take into account the interaction between the two: since the output of the model is not used by the DLS when the supervisor activates the fail safe procedure ($u(i) \geq t$), it does not make any sense to evaluate the performance of the model in such scenarios. For this reason, we propose a new approach for the joint assessment of model and supervisor, which we call \textit{supervised metrics}. In the next two sections we first summarize the state of the art metrics for the separate, individual assessment of model and supervisor, followed by a description of our proposal of a new joint assessment approach.


\subsection{Existing Metrics for the Individual Assessment of Model and Supervisor}

There are well established metrics for the individual assessment of  performance of a model $m$. These are  based on some \emph{objective function} $obj(I,m)$, such as \emph{accuracy (ACC)} (for classifiers) or \emph{mean-squared error (MSE)} (for regression models), computed on a test dataset $I$.\footnote{We assume that $obj$ has to be maximized. The extension
to an  $obj$ that should be minimized is straightforward. Also note that $obj$ does not have to be the same function used to optimize the DNN during training.}

Classically, the supervisor's performance would be assessed individually using  performance metrics designed for binary classifiers. For a given $t$, the available metrics include \emph{true positive rate (TPR)}, \emph{false positive rate (FPR)}, \emph{true negative rate (TNR)} and \emph{false negative rate (FNR)}, \emph{$F_1$} score and \emph{accuracy (ACC)}. 
To use these metrics with regression problems, an \emph{acceptable imprecision} would have to be defined, allowing to divide inputs into benign (negative cases) and malicious (positive cases).
Alternatively, the effect of the predictions on the overall DLS system could be monitored to only treat inputs leading to system failures as malicious ones~\cite{Stocco2020}.

There are also existing, classical metrics to assess a binary classifier independently of the threshold $t$.
For instance, the \emph{average precision score (AVGPR)} computes
the average of the precision values obtained when varying $t$, weighted by the recall measured at $t$. 
Another popular threshold independent metric is the \emph{area under the receiver operating characteristic curve (AUROC)}\cite{Stocco2020, Berend2020, Michelmore2018}. 
When individually assessing the performance of a supervisor, AVGPR should be preferred over AUROC as, amongst other advantages \cite{Davis2006}, it is better suited for the unbalanced datasets~\cite{Saito2015} typically observed during malicious input detection.
Threshold independent analysis of the supervisor in a regression problem is  straightforward and can be done using point-biserial correlation between the quantified uncertainty and the observed prediction error, given some objective function (e.g. MSE). 

Independent analysis of model's performance, considered in isolation without supervision, and of the supervisor's performance, again in isolation, results in measurements that do not capture the overall goal of the interaction between supervisor and model under supervision: ensuring high model performance on the samples considered safe by the supervisor, while keeping the amount of samples considered unsafe as small as possible. 
To capture such goal, we propose novel metrics for joint model and uncertainty quantification assessment in a supervised DLS.

\subsection{Supervised Metrics for the Joint Assessment of Model and Supervisor}
When considering model and supervisor jointly, we can still use the objective function used to assess the model in isolation, but  we evaluate it in a supervised context:
given a test set $I$ and an objective function $obj(I,m)$ for model $m$ with uncertainty quantifier $u$ and supervisor threshold $t$, the \textit{supervised objective function} $\overline{obj}(I,m)$ is defined as:
$$
\overline{obj}(m,I) = obj(\{i \mid i \in I \text{ and } u(i) < t \})
$$
i.e., the objective is applied only to the subset of inputs which is accepted by the supervisor.
By decreasing $t$, assuming  that the cardinality of the resulting subset of inputs remains big enough to calculate a statistically significant $\overline{obj}(I)$, we may generally get higher values of the supervised objective function $\overline{obj}(I)$.
However, such high values of $\overline{obj}(I)$ are likely associated with a high false alarm rate of the supervisor.
Thus, any $\overline{obj}(I)$ should always be regarded in conjunction with the acceptance rate $\Delta_u(I)$ of the supervisor:
$$
\Delta_u(I) = \frac{\mid \{i \mid i \in I \text{ and } u(i) < t \} \mid }{\mid I \mid}
$$

Similar to the popular $F1$ score, the following combination of these two metrics allows to capture the effectiveness of the collaboration between supervised model and supervisor:
\begin{termdefinition}[S-Score]
The $S_1$-Score measures the harmonic mean of a supervised objective function, normalized between zero and one, and the supervisors acceptance rate as
$$
S_1(m,u,I) = \frac{2}{\frac{obj^+}{\overline{obj}(I,m) - obj^-} + \Delta_u(I)^{-1}}
$$
\end{termdefinition}
where $obj^-$ and $obj^+$ are the lower and upper bounds used for normalization of the objective function.
For classifiers, if  accuracy is the objective function,  $obj^- := 0$ and $obj^+ := 1$. For regression problems, or more generally for unbounded objective functions, $obj^-$ and $obj^+$ have to be estimated empirically (e.g., based on the empirical distribution of the objective function values), independently from $m$ and $u$.

The $S_1$ scores weights $\Delta_u(I)$ and $\overline{obj}(m,I)$ equally. 
Equivalent to the popular $F_1$ score, other $S_\beta$ scores can be used, where $\beta > 0$ is the weighting parameter~\cite{Rijsbergen1979}:

$$
S_\beta(m,u,I) = (1 + \beta^2)\cdot \frac{\frac{\overline{obj}(I,m) - obj^-}{obj^+} \cdot \Delta_u(I)}{(\beta^2 \cdot \frac{\overline{obj}(I,m) - obj^-}{obj^+}) + \Delta_u(I)}
$$

\section{Case Studies}


\begingroup
\setlength{\tabcolsep}{5.5pt} 
\renewcommand{\arraystretch}{0.9} 

\newcommand{\vertimodeltype}[1]{\begin{tabular}{@{}c@{}}\rotatebox[origin=c]{90}{\parbox{1cm}{\centering #1}}\end{tabular}}
\def \spacebetweenbigtrows {\vspace{0.15cm}}

\begin{table*}[t]
\scriptsize
\begin{tabular}{lllrrrrrrrrrrrrrrrrrr}
\toprule
                             &                              &            & \multicolumn{9}{c}{Nominal (regular test data)}                                                                                                                                                                                                                                    & \multicolumn{9}{c}{Out of Distribution (corrupted test data)}\\
\cmidrule(r){4-12}
\cmidrule(r){13-21}
                             &                              &            &  & \multicolumn{4}{c}{$\epsilon$ = 0.01}                                                                        & \multicolumn{4}{c}{$\epsilon$ = 0.1}                                                                         &  & \multicolumn{4}{c}{$\epsilon$ = 0.01}                                                                        & \multicolumn{4}{c}{$\epsilon$ = 0.1}                                                                         \\
\cmidrule(r){5-8}
\cmidrule(r){9-12}
\cmidrule(r){14-17}
\cmidrule(r){18-21}

& \multicolumn{2}{c}{Technique}                & \multicolumn{1}{r}{$\scriptstyle ACC$}        & S-C & \multicolumn{1}{r}{$\overline{\scriptstyle ACC}$} & \multicolumn{1}{r}{$\Delta_u$} & \multicolumn{1}{r}{$S_1$} & S-C & \multicolumn{1}{r}{$\overline{\scriptstyle ACC}$} & \multicolumn{1}{r}{$\Delta_u$} & \multicolumn{1}{r}{$S_1$} & \multicolumn{1}{r}{$\scriptstyle ACC$}        & S-C & \multicolumn{1}{r}{$\overline{\scriptstyle ACC}$} & \multicolumn{1}{r}{$\Delta_u$} & \multicolumn{1}{r}{$S_1$} & S-C & \multicolumn{1}{r}{$\overline{\scriptstyle ACC}$} & \multicolumn{1}{r}{$\Delta_u$} & \multicolumn{1}{r}{$S_1$} \\
\midrule
\multirow{11}{*}{\vertimodeltype{Cifar10}}    & \multirow{3}{*}{\vertimodeltype{Point Pred.}} & SM         & 0.82                             & -0.67    & 0.83                                 & 0.97                           & 0.90                      & -0.67    & 0.90                                 & 0.80                           & 0.85                      & 0.82                             & -0.47    & 0.83                                 & 0.98                           & 0.90                      & -0.52    & 0.89                                 & 0.81                           & 0.85                      \\
                             &                              & PCS        & 0.82                             & -0.71    & 0.83                                 & 0.97                           & 0.90                      & -0.70    & 0.90                                 & 0.80                           & 0.85                      & 0.82                             & -0.49    & 0.83                                 & 0.97                           & 0.90                      & -0.54    & 0.89                                 & 0.81                           & 0.85                      \\
                             &                              & SME        & 0.82                             & -0.64    & 0.84                                 & 0.97                           & 0.90                      & -0.61    & 0.91                                 & 0.80                           & 0.85                      & 0.82                             & -0.50    & 0.83                                 & 0.97                           & 0.90                      & -0.54    & 0.89                                 & 0.80                           & 0.84                      \\
                             & \multirow{4}{*}{\vertimodeltype{MC- Dropout}}  & VR         & 0.82                             & -0.68    & 0.84                                 & 0.98                           & 0.90                      & -0.68    & 0.90                                 & 0.82                           & 0.86                      & 0.82                             & -0.63    & 0.83                                 & 0.98                           & 0.90                      & -0.62    & 0.89                                 & 0.82                           & 0.85                      \\
                             &                              & PE         & 0.82                             & -0.70    & 0.84                                 & 0.97                           & 0.90                      & -0.63    & 0.90                                 & 0.82                           & 0.86                      & 0.82                             & -0.59    & 0.83                                 & 0.97                           & 0.90                      & -0.43    & 0.88                                 & 0.83                           & 0.85                      \\
                             &                              & MI         & 0.82                             & -0.76    & 0.84                                 & 0.97                           & 0.90                      & -0.64    & 0.89                                 & 0.82                           & 0.86                      & 0.82                             & -0.57    & 0.83                                 & 0.98                           & 0.90                      & -0.54    & 0.88                                 & 0.83                           & 0.86                      \\
                             &                              & MS         & 0.82                             & -0.66    & 0.84                                 & 0.97                           & 0.90                      & -0.67    & 0.91                                 & 0.81                           & 0.86                      & 0.82                             & -0.56    & 0.83                                 & 0.98                           & 0.90                      & -0.52    & 0.89                                 & 0.81                           & 0.85                      \\
                             & \multirow{4}{*}{\vertimodeltype{Ensem- ble}}    & VR         & 0.86                             & -0.66    & 0.87                                 & 0.97                           & 0.92                      & -0.78    & 0.93                                 & 0.83                           & 0.88                      & 0.87                             & -0.67    & 0.87                                 & 0.98                           & 0.92                      & -0.74    & 0.93                                 & 0.83                           & 0.87                      \\
                             &                              & PE         & 0.86                             & -0.78    & 0.87                                 & 0.98                           & 0.92                      & -0.81    & 0.92                                 & 0.84                           & 0.88                      & 0.87                             & -0.69    & 0.87                                 & 0.98                           & 0.92                      & -0.63    & 0.91                                 & 0.85                           & 0.88                      \\
                             &                              & MI         & 0.86                             & -0.78    & 0.87                                 & 0.98                           & 0.92                      & -0.83    & 0.92                                 & 0.84                           & 0.88                      & 0.87                             & -0.68    & 0.87                                 & 0.98                           & 0.92                      & -0.72    & 0.91                                 & 0.84                           & 0.87                      \\
                             &                              & MS         & 0.86                             & -0.73    & 0.87                                 & 0.97                           & 0.92                      & -0.80    & 0.94                                 & 0.83                           & 0.88                      & 0.86                             & -0.67    & 0.87                                 & 0.98                           & 0.92                      & -0.64    & 0.93                                 & 0.83                           & 0.88\spacebetweenbigtrows                      \\
\multirow{11}{*}{\vertimodeltype{Mnist}}      & \multirow{3}{*}{\vertimodeltype{Point Pred.}} & SM         & 0.96                             & -0.80    & 0.97                                 & 0.99                           & 0.98                      & -0.89    & 0.99                                 & 0.88                           & 0.93                      & 0.74                             & -0.13    & 0.84                                 & 0.82                           & 0.83                      & -0.88    & 0.96                                 & 0.48                           & 0.64                      \\
                             &                              & PCS        & 0.96                             & -0.79    & 0.97                                 & 0.99                           & 0.98                      & -0.86    & 0.99                                 & 0.88                           & 0.93                      & 0.74                             & -0.32    & 0.80                                 & 0.88                           & 0.84                      & -0.87    & 0.96                                 & 0.49                           & 0.65                      \\
                             &                              & SME        & 0.96                             & -0.90    & 0.97                                 & 0.99                           & 0.98                      & -0.92    & 1.00                                 & 0.88                           & 0.93                      & 0.74                             & -0.20    & 0.85                                 & 0.79                           & 0.82                      & -0.90    & 0.96                                 & 0.47                           & 0.63                      \\
                             & \multirow{4}{*}{\vertimodeltype{MC- Dropout}}  & VR         & 0.97                             & -0.63    & 0.97                                 & 0.98                           & 0.98                      & -0.77    & 0.99                                 & 0.87                           & 0.93                      & 0.74                             & -0.15    & 0.82                                 & 0.85                           & 0.84                      & -0.64    & 0.95                                 & 0.51                           & 0.67                      \\
                             &                              & PE         & 0.97                             & -0.73    & 0.97                                 & 0.99                           & 0.98                      & -0.81    & 1.00                                 & 0.87                           & 0.93                      & 0.74                             & -0.10    & 0.85                                 & 0.79                           & 0.82                      & -0.81    & 0.97                                 & 0.45                           & 0.62                      \\
                             &                              & MI         & 0.97                             & -0.72    & 0.97                                 & 0.98                           & 0.98                      & -0.87    & 0.99                                 & 0.87                           & 0.93                      & 0.74                             & -0.15    & 0.79                                 & 0.88                           & 0.83                      & -0.61    & 0.91                                 & 0.55                           & 0.69                      \\
                             &                              & MS         & 0.96                             & -0.71    & 0.97                                 & 0.99                           & 0.98                      & -0.84    & 1.00                                 & 0.88                           & 0.93                      & 0.74                             & -0.03    & 0.84                                 & 0.82                           & 0.83                      & -0.76    & 0.96                                 & 0.47                           & 0.64                      \\
                             & \multirow{4}{*}{\vertimodeltype{Ensem- ble}}    & VR         & 0.97                             & -0.74    & 0.98                                 & 0.98                           & 0.98                      & -0.81    & 0.98                                 & 0.95                           & 0.97                      & 0.74                             & -0.30    & 0.86                                 & 0.78                           & 0.82                      & -0.75    & 0.91                                 & 0.65                           & 0.76                      \\
                             &                              & PE         & 0.97                             & -0.89    & 0.97                                 & 0.99                           & 0.98                      & -0.93    & 0.99                                 & 0.88                           & 0.93                      & 0.74                             & -0.34    & 0.86                                 & 0.77                           & 0.81                      & -0.89    & 0.97                                 & 0.45                           & 0.61                      \\
                             &                              & MI         & 0.97                             & -0.82    & 0.97                                 & 0.98                           & 0.98                      & -0.87    & 1.00                                 & 0.87                           & 0.93                      & 0.74                             & -0.45    & 0.87                                 & 0.71                           & 0.78                      & -0.86    & 0.98                                 & 0.42                           & 0.59                      \\
                             &                              & MS         & 0.97                             & -0.87    & 0.97                                 & 0.98                           & 0.98                      & -0.89    & 1.00                                 & 0.88                           & 0.93                      & 0.75                             & -0.48    & 0.86                                 & 0.79                           & 0.82                      & -0.91    & 0.96                                 & 0.46                           & 0.63\spacebetweenbigtrows                      \\
\multirow{11}{*}{\vertimodeltype{Traffic}}    & \multirow{3}{*}{\vertimodeltype{Point Pred.}} & SM         & 0.81                             & -0.02    & 0.90                                 & 0.89                           & 0.89                      & -0.27    & 0.97                                 & 0.75                           & 0.85                      & 0.79                             & 0.02     & 0.90                                 & 0.85                           & 0.88                      & -0.16    & 0.97                                 & 0.70                           & 0.81                      \\
                             &                              & PCS        & 0.81                             & 0.04     & 0.90                                 & 0.89                           & 0.89                      & -0.21    & 0.97                                 & 0.75                           & 0.84                      & 0.79                             & 0.03     & 0.90                                 & 0.85                           & 0.88                      & -0.20    & 0.97                                 & 0.70                           & 0.81                      \\
                             &                              & SME        & 0.81                             & -0.09    & 0.90                                 & 0.89                           & 0.89                      & -0.38    & 0.98                                 & 0.72                           & 0.83                      & 0.79                             & 0.00     & 0.90                                 & 0.86                           & 0.88                      & -0.36    & 0.98                                 & 0.66                           & 0.79                      \\
                             & \multirow{4}{*}{\vertimodeltype{MC- Dropout}}  & VR         & 0.81                             & -0.24    & 0.91                                 & 0.87                           & 0.89                      & -0.42    & 0.98                                 & 0.70                           & 0.82                      & 0.79                             & -0.40    & 0.91                                 & 0.84                           & 0.88                      & -0.42    & 0.98                                 & 0.66                           & 0.79                      \\
                             &                              & PE         & 0.81                             & -0.18    & 0.90                                 & 0.89                           & 0.89                      & -0.40    & 0.98                                 & 0.70                           & 0.81                      & 0.79                             & -0.14    & 0.89                                 & 0.86                           & 0.88                      & -0.42    & 0.98                                 & 0.65                           & 0.78                      \\
                             &                              & MI         & 0.81                             & -0.23    & 0.90                                 & 0.88                           & 0.89                      & -0.44    & 0.98                                 & 0.70                           & 0.81                      & 0.79                             & -0.37    & 0.89                                 & 0.87                           & 0.88                      & -0.44    & 0.98                                 & 0.65                           & 0.78                      \\
                             &                              & MS         & 0.81                             & -0.09    & 0.91                                 & 0.87                           & 0.89                      & -0.38    & 0.98                                 & 0.70                           & 0.81                      & 0.79                             & -0.19    & 0.91                                 & 0.84                           & 0.88                      & -0.37    & 0.98                                 & 0.65                           & 0.78                      \\
                             & \multirow{4}{*}{\vertimodeltype{Ensem- ble}}    & VR         & 0.81                             & -0.56    & 0.94                                 & 0.84                           & 0.89                      & -0.57    & 0.97                                 & 0.76                           & 0.86                      & 0.80                             & -0.47    & 0.93                                 & 0.82                           & 0.87                      & -0.64    & 0.97                                 & 0.73                           & 0.83                      \\
                             &                              & PE         & 0.81                             & -0.46    & 0.92                                 & 0.85                           & 0.89                      & -0.47    & 0.98                                 & 0.69                           & 0.81                      & 0.80                             & -0.55    & 0.93                                 & 0.82                           & 0.87                      & -0.51    & 0.99                                 & 0.63                           & 0.77                      \\
                             &                              & MI         & 0.81                             & -0.59    & 0.94                                 & 0.84                           & 0.89                      & -0.46    & 0.98                                 & 0.69                           & 0.81                      & 0.80                             & -0.64    & 0.94                                 & 0.82                           & 0.87                      & -0.56    & 0.99                                 & 0.63                           & 0.77                      \\
                             &                              & MS         & 0.81                             & -0.47    & 0.93                                 & 0.85                           & 0.89                      & -0.48    & 0.98                                 & 0.69                           & 0.81                      & 0.80                             & -0.45    & 0.93                                 & 0.82                           & 0.87                      & -0.52    & 0.99                                 & 0.63                           & 0.77\spacebetweenbigtrows                      \\
\multirow{7}{*}{\vertimodeltype{Imagent}} & \multirow{3}{*}{\vertimodeltype{Point Pred.}} & SM         & 0.74                             & n.a.     & 0.77                                 & 0.95                           & 0.85                      & n.a.     & 0.84                                 & 0.79                           & 0.81                      & 0.50                             & n.a.     & 0.61                                 & 0.79                           & 0.69                      & n.a.     & 0.75                                 & 0.54                           & 0.63                      \\
                             &                              & PCS        & 0.74                             & n.a.     & 0.76                                 & 0.97                           & 0.85                      & n.a.     & 0.84                                 & 0.79                           & 0.81                      & 0.50                             & n.a.     & 0.55                                 & 0.89                           & 0.68                      & n.a.     & 0.72                                 & 0.57                           & 0.64                      \\
                             &                              & SME        & 0.74                             & n.a.     & 0.76                                 & 0.96                           & 0.85                      & n.a.     & 0.83                                 & 0.80                           & 0.81                      & 0.50                             & n.a.     & 0.60                                 & 0.80                           & 0.68                      & n.a.     & 0.73                                 & 0.55                           & 0.63                      \\
                             & \multirow{4}{*}{\vertimodeltype{MC- Dropout}}  & VR         & 0.74                             & -0.37    & 0.76                                 & 0.96                           & 0.85                      & -0.73    & 0.84                                 & 0.78                           & 0.81                      & 0.50                             & -0.69    & 0.56                                 & 0.87                           & 0.68                      & -0.84    & 0.71                                 & 0.57                           & 0.63                      \\
                             &                              & PE         & 0.74                             & -0.12    & 0.76                                 & 0.96                           & 0.85                      & -0.43    & 0.83                                 & 0.79                           & 0.81                      & 0.50                             & -0.48    & 0.61                                 & 0.78                           & 0.69                      & -0.51    & 0.75                                 & 0.53                           & 0.62                      \\
                             &                              & MI         & 0.74                             & -0.39    & 0.75                                 & 0.98                           & 0.85                      & -0.52    & 0.82                                 & 0.81                           & 0.81                      & 0.50                             & -0.52    & 0.52                                 & 0.93                           & 0.67                      & -0.76    & 0.67                                 & 0.58                           & 0.62                      \\
                             &                              & MS         & 0.74                             & -0.07    & 0.77                                 & 0.96                           & 0.85                      & -0.17    & 0.85                                 & 0.78                           & 0.81                      & 0.50                             & -0.40    & 0.61                                 & 0.79                           & 0.69                      & -0.37    & 0.76                                 & 0.52                           & 0.62                     
 \\

                             \bottomrule
\end{tabular}
\caption{Overview of supervision capabilities of different techniques, i.e., model types and quantifier combinations for different thresholds.}
\label{tab:big_table}
\end{table*}

\endgroup
\label{sec:case_studies}
We assess the uncertainty quantification capabilities of point predictors, deep ensemble and MC dropout using different quantifiers. 
We intentionally focus our study on uncertainty quantifiers which can be applied to traditional and widely used DNN architectures, and we exclude those implementing the pure Bayesian form of uncertainty estimation, since they require the adoption of dedicated architectures where the network weights encode a probability distribution, not just a scalar value.
This restriction, in combination with \tool, allows developers to measure uncertainty at minimal effort,
given a traditional DNN.
The \textit{goal} of our empirical evaluation is to assess the usefulness of the uncertainty quantifiers supported by \tool when used as supervisors, as well as to collect lessons learned that practitioners can follow when applying \tool to their DNNs.

\subsection{Research Questions}

We consider the following research questions:

\noindent
\textbf{RQ\textsubscript{1} (effectiveness):}
\textit{How effective are  supervisors at increasing the supervised model's performance?}

This is the key research question of our empirical study, since the main hypothesis behind supervisors is that they can prevent usage of a model when its performance is predicted to be low. Hence, we expect an increase of the supervised model's performance $\overline{obj}$ as compared to the unsupervised one $obj$.

\noindent
\textbf{RQ\textsubscript{2} (comparison):}
\textit{Is there a supervisor and quantifier type which yield optimal performance across  subjects and across alternative choices of the uncertainty threshold?}

We consider three types of uncertainty estimators, Point Predictors, MC-Dropout and Ensemble, and several uncertainty quantifiers, respectively [SM, PCS, SME], [VR, PE, MI, MS], [VR, PE, MI, MS] (see Section~\ref{sec:approaches}). We want to investigate whether any combination of estimator type and quantifier dominates all the others in terms of $S_1$-score. To investigate how performance changes with the uncertainty threshold $t$, we consider different acceptable rates $\epsilon$ of false positives on the nominal data and we compute the threshold $t$ that ensures such FPR on the validation set, so that we can compare alternative estimators/quantifiers at equal FPR on the validation set of the nominal data.

\noindent
\textbf{RQ\textsubscript{3} (sample size):}
\textit{How many samples are required in stochastic and ensemble models to get reliable uncertainty quantification?}

Since the main cost associated with the usage of MC-Dropout and Ensemble is the collection of multiple samples for each individual prediction, we want to understand what is the minimum sample size that ensures good performance of each different supervisor. In particular, we study the convergence of  supervised accuracy to its asymptotic value as the sample size is increased.

\noindent
\textbf{RQ\textsubscript{4} (sensitivity):}
\textit{How sensitive are supervisors to changes in their main hyperparameters?}

With this research question we want to understand whether the choice of hyperparameters is critical to get optimal performance, or  on the contrary if they can be chosen in the neighbourhood of the optimal choice with minor impact on the resulting performance of the supervisor.
For Point Predictors, we consider the number of training epochs as the main hyperparameter;
for MC-Dropout, the number of training epochs and the number of samples; for Ensemble, the number of training epochs and the number of atomic models. We measure the standard deviation of the supervised objective function (e.g., supervised accuracy) in the neighbourhood of each hyperparameter choice, so as to identify the regions where such standard deviation is low.

\subsection{Subjects}

We use the following classification problems as case study subjects, aiming to increase diversity and practical relevance. 
\begin{description}
    \item [Mnist\cite{LeCun1998}] Classification of hand-written digits, formatted as small grayscale images. 
    This is the most popular dataset in machine learning testing~\cite{Riccio2020}, and a relatively easy problem, where even simple models achieve high accuracy. 
    We took the DNN architecture from a Keras tutorial~\cite{KerasMnistModel}.
    
    \item [Cifar10\cite{Krizhevsky2009}] Classification of colored images into ten different classes. It is also very popular in DLS testing~\cite{Riccio2020} and it represents a more challenging task than Mnist.
    We use the model architecture proposed in the Brownlees Cifar10 tutorial~\cite{BrownleeCifarModel}.
    
    \item [Traffic~\cite{Serna2018}] Classification of images of European traffic signs~\cite{Segvic2010, Bonaci2011, Mathias2013, Timofte2014, Belaroussi2010, Stallkamp2012, Grigorescu2003, Larsson2011}.
    The different sources the data was collected from, combined with the fact that the dataset is unbalanced and many images are of bad quality, reflect a quite realistic, high-uncertainty setup. Since traffic sign recognition is a core component of self-driving cars, this is also a very interesting case study from the software and system engineering point of view.
    The model architecture we use was proposed alongside the release of this dataset~\cite{Serna2018}.
    
    \item [Imagenet~\cite{Deng2009} (Pretrained)] Image classification problem with as many as 1,000 classes.
    We use  eight pre-trained \emph{Efficientnet} models~\cite{Tan2019}.
    As for this subject we rely on pre-trained models (which include dropout layers), we can test them only 
    as MC-Dropout and Point Predictor models, but not as Ensembles.
\end{description}

\subsection{Experimental Setup}
Except for the pre-trained ones, models were trained for 200 epochs. 
After every epoch, we assessed the models' performance on both a nominal and an out-of-distribution (OOD) dataset, for every quantifier.
To do so, we used Mnist-c~\cite{Mu2019} as OOD test set for Mnist and the color-image transformations
proposed by Hendrycks~\cite{Hendrycks2018} to generate OOD samples for the other subjects.
We used three different thresholds, calculated on the nominal validation set to ensure the lowest possible FPR above $\epsilon$, with $\epsilon\in\{.01,.5,.1\}$ respectively.
To measure the sensitivity to the number of samples, quantifiers of deep ensembles were assessed with every number of atomic models between 2 and 50,
Similarly, MC-Dropout was assessed on every number of samples between 2 and 100.

Counting atomic models individually, this procedure required the training of 153 DNNs,
and the calculation of 2'121'600 DNN predictions\footnote{Predictions were cached, such that for the evaluation of different sample sizes and different numbers
of atomic models, previous predictions could be re-used.}.
Due to the high workload, the training and prediction processes were distributed on three different workstations using Windows or Ubuntu and four different GPUs (one workstation had two GPUs). 
\tool was used for training, prediction and uncertainty quantification.
\footnote{Replication package available at:\\ \href{https://github.com/testingautomated-usi/repli-icst2021-uncertainty}{github.com/testingautomated-usi/repli-icst2021-uncertainty}}
\subsection{Results}

We organize the analysis of the results obtained in our experiments by research question.

\subsubsection{RQ1 (Effectiveness)}

An overview of our results is provided in \autoref{tab:big_table}. 
Due to space constraints, the results for $\epsilon=0.05$ are omitted in the table and the values for the 8 different Imagenet models are averaged. 
The full set of results can be found in the online replication package.

Our results suggest that all supervisors lead to supervised accuracies $\overline{ACC}$ which are at least as high, but typically much higher than the accuracy $ACC$ of the unsupervised model.
Thus,  supervisors are effective. 
The effectiveness is particularly strong on the OOD datasets:
For example, on Mnist, where the unsupervised point predictor has an accuracy of 74\%, 
the supervised accuracy at $\epsilon=0.1$, 
is above 95\% with most supervisors. 
In other words, a DLS using an unsupervised model will experience six times more faulty predictions than the unsupervised one.
Also notable are the results on the nominal Mnist dataset, where even simple supervisors based on  point predictors turn an unsupervised accuracy of 96\% (at $\epsilon = 0.1$) into that of a nearly perfect predictor 
while still accepting around 88\% of the inputs.

\begin{tcolorbox}
\textbf{Summary (RQ1)}: \textit{All tested supervisors are, in general, effective at increasing the supervised accuracy compared to the unsupervised accuracy.} 
\end{tcolorbox}

\subsubsection{RQ2 (Comparison)}
\newcommand{\verti}[1]{\begin{tabular}{@{}c@{}}\rotatebox[origin=c]{90}{\parbox{1cm}{\centering #1}}\end{tabular}}
\def \spacebetweenrankrows {0.1cm}
\begin{table}[t]
\scriptsize
\centering
\begin{tabular}{@{}llcccccccc@{}}
\toprule
                                      &             &  \multicolumn{3}{c}{Per Subject}                                                                                          &&   \multicolumn{2}{c}{Overall}             \\
                                      &              & mnist                                    & cifar10                                  & traffic                                   &&  Trained                                & Pre-Trained                                                   \\
\multicolumn{2}{c}{Technique}                       & N=6                                & N=6                                & N=6                                &&  N=18                                 & N=48           \\ 
\midrule
\multirow{3}{*}{\verti{Point Pred.}}   & SM     & 3.33     & 9.67     & \textbf{2.42}     &&  5.14     & 3.56             \\
                                      & PCS         & \textbf{2.67}             & 9.83             & 3.17             &&  5.22             & 3.5                       \\
                                      & SME     & 5.67 & 10.0 & 2.92 &&  6.19 & 4.42             \vspace{\spacebetweenrankrows}    \\
                                 
\multirow{4}{*}{\verti{MC- Dropout}}      & VR  & 5.83       & 5.83       & 5.67       &&  5.78       & 3.55             \\
                                                   & PE  & 7.0    & 6.67    & 5.5    &&  6.39    & 4.5             \\
                                                   & MI & 6.67       & 6.83       & 5.75       &&  6.42       & 5.12             \\
                                                   & MS & 5.33         & 7.17         & 6.58         &&  6.36         & \textbf{3.34}  \vspace{\spacebetweenrankrows}    \\
\multirow{4}{*}{\verti{Ensem- ble}}       & VR  & 4.17         & 3.17         & 4.5         &&  \textbf{3.94}        & n.a.                                                 \\
                                                   & PE  & 8.17      & \textbf{1.33}     & 10.25      &&  6.58      & n.a.                                                 \\
                                                   & MI & 10.17         & 3.0         & 10.25         &&  7.81         & n.a.                                                 \\
                                                   & MS & 7.0           & 2.5           & 9.0           &&  6.17           & n.a. \\
                                                   \bottomrule
\end{tabular}
\caption{Rank-Order Analysis: $S_1$-Score ranks, averaged over $\epsilon\in\{0.01, 0.05, 0.1\}$, nominal and OOD datasets}
\label{tab:rank_order}
\end{table}

If we look at the $S_1$-Score in \autoref{tab:big_table}, it is apparent that there is no uncertainty quantifier which outperforms all the other ones on every subject/dataset and for every threshold ($\epsilon$). 
To allow for an overall comparison of the quantifiers, we computed the average ranks of the quantifiers  when ordered by $S_1$-Score. Results are shown in \autoref{tab:rank_order}, where $N$ indicates the number of data points on which average ranks have been computed. While there is no absolute dominant supervisor, i.e., the ideal choice of supervisor remains problem dependent,
Ensembles can be considered the overall best performing supervisors (in line with existing literature\cite{Ovadia2019}), because when they do not have the lowest rank, they still have quite low rank values. Actually, 
even without supervision Ensembles often achieve higher accuracy than Point Predictors and MC-Dropout based models.
Interestingly, in most cases the sophisticated quantifiers PE and MI, grounded on information theory,
do not perform better and often perform  worse than the simpler quantifiers VR and AS. 
Despite the theoretical disadvantages of Point Predictors, in our experiments
this simple approach outperformed the theoretically well founded MC-Dropout. So, in practical cases as those considered in our experiments, Point Predictors may represent a good trade-off between performance and computational cost.

\begin{tcolorbox}
\textbf{Summary (RQ2)}: \textit{There is no dominant supervisor, i.e., no supervisor which performs best for every test subject, data source and threshold. Ensembles are ranked generally well across subjects and thresholds, while Point Predictors offer a valuable trade off between performance and execution cost.} 
\end{tcolorbox}

\subsubsection{RQ3 (Sample size)}
\def \sssubfigurewidth {.625\linewidth}
\def \ssimgwidth {\linewidth}

\begin{figure}
\centering
\begin{subfigure}{\sssubfigurewidth}
  \centering
  \includegraphics[width=\ssimgwidth]{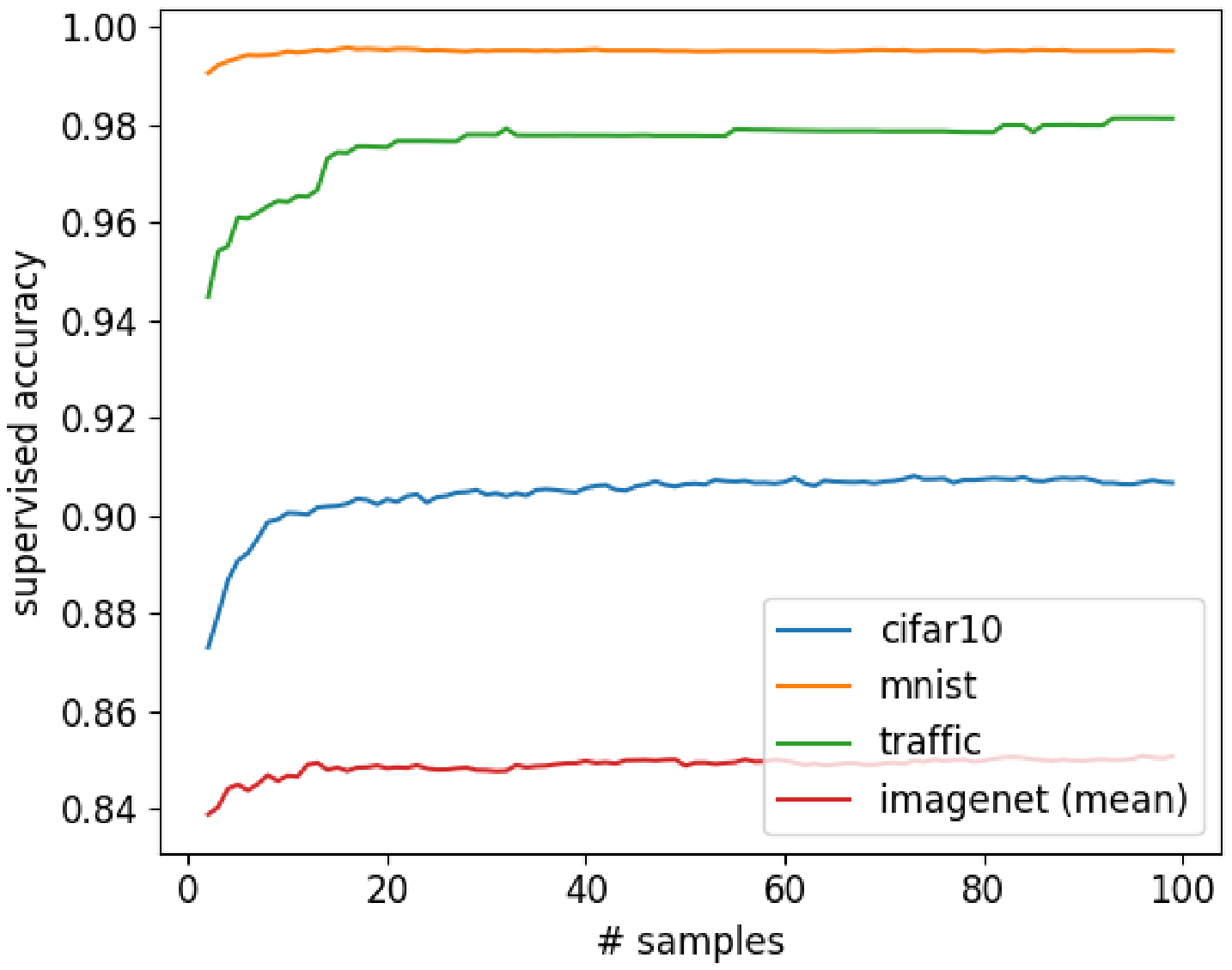}
  \caption{}
  \label{fig:ss_influence_mcdropout}
\end{subfigure}%
\hfill
\begin{subfigure}{0.375\linewidth}
  \centering
  \includegraphics[width=\ssimgwidth]{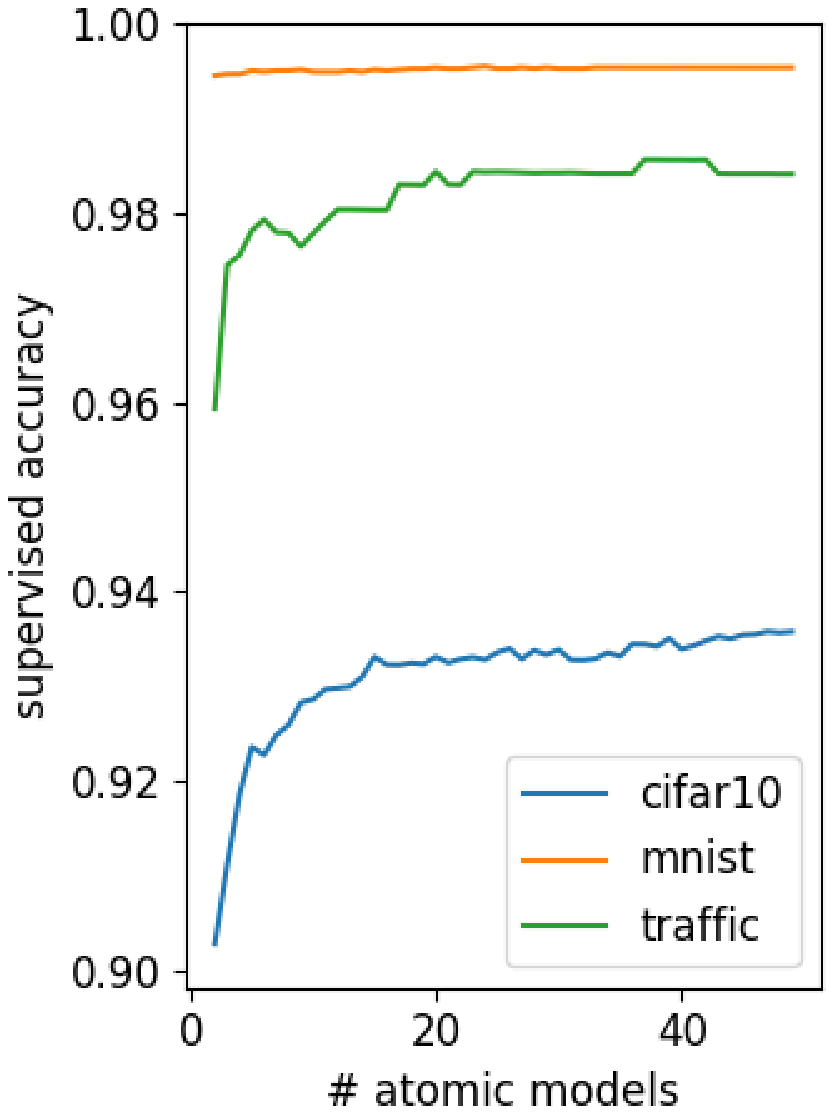}
  \caption{}
  \label{fig:ss_influence_ensemble}
\end{subfigure}

\caption{Influence of sample size in MC-Dropout (\subref{fig:ss_influence_mcdropout}) and number of atomic models in Ensemble (\subref{fig:ss_influence_ensemble}) on supervised accuracy $\overline{acc}$;
values taken with $\epsilon=0.1$ and MS quantifier on nominal dataset}
\label{fig:ss_influence}
\end{figure}
We find that for both MC-Dropout and Ensembles the relatively low number of 20 samples
is already sufficient to get a similar supervised accuracy as with a much higher number of samples.
This is shown in \autoref{fig:ss_influence} for the MS quantifier and $\epsilon = 0.1$. 
20 samples, while still higher than what's recommended in related literature~\cite{Ovadia2019},
is probably small enough to be used in practice in many applications.
The other quantifiers behave similarly, with one notable exception, which is the VR quantifier: VR can only take a finite set of discrete values, which can be easily shown to be equal to the number of samples. 
Hence, a low sample size makes it impossible to set thresholds well fit to the target $\epsilon$, because thresholds are correspondingly also discrete and limited to the number of samples. So, to achieve the target FPR $\epsilon$ precisely, we might need substantially more than 20 samples.

\begin{tcolorbox}
\textbf{Summary (RQ3)}: \textit{A few (\texttildelow20) samples are  enough to get good supervision results with most quantifiers (VR represents an exception, due to the discretization of the values it can take).} 
\end{tcolorbox}

\subsubsection{RQ4 (Sensitivity)}
\def \hmsubfigurewidth {.31\linewidth}

\def \hmimgwidth {4cm}
\def \hmimgheight {4.45cm}

\begin{figure}
\centering
\begin{subfigure}{\hmsubfigurewidth}
  \centering
  \includegraphics[
  height=\hmimgheight
  ]{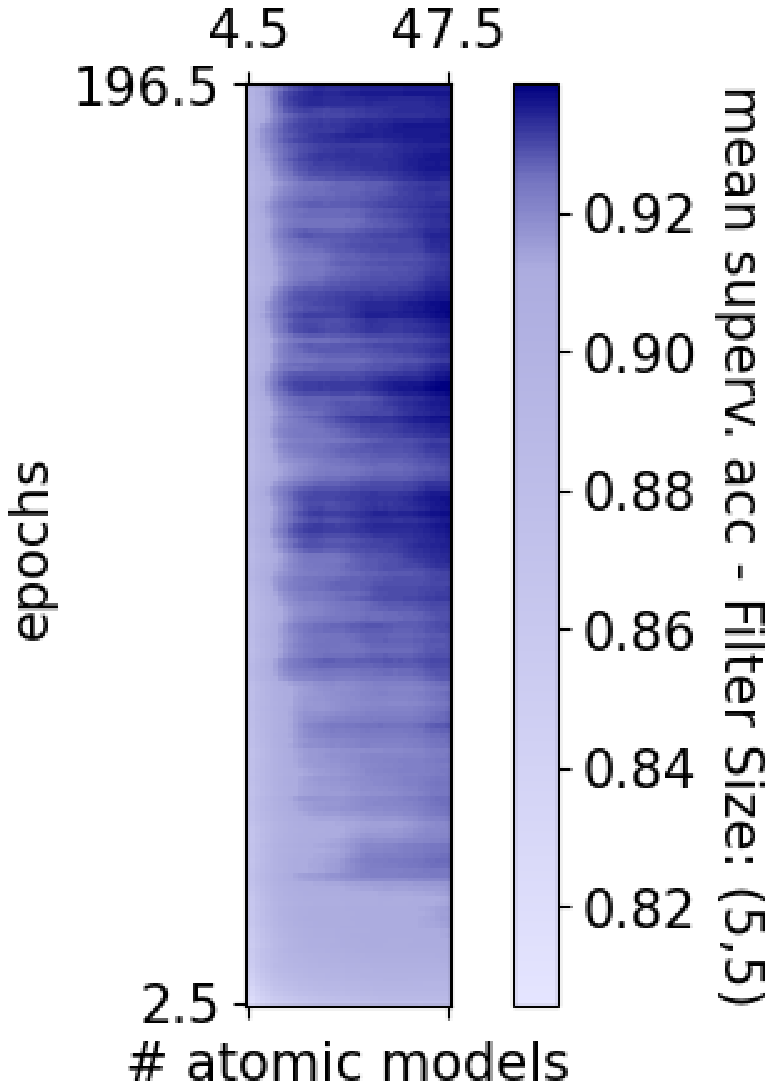}
  \caption{}
  \label{fig:heatmap_mean}
\end{subfigure}%
\hspace{1cm}
\begin{subfigure}{\hmsubfigurewidth}
   \centering
  \includegraphics[
  height=\hmimgheight
  ]{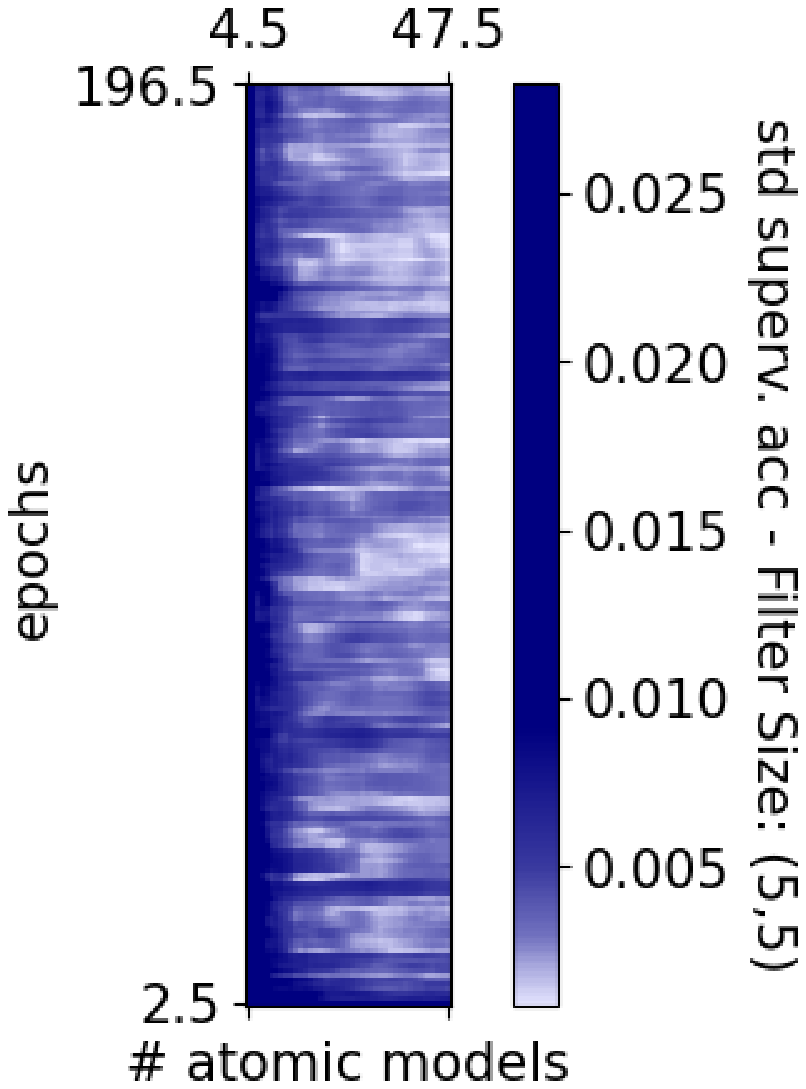}
  \caption{}
  \label{fig:heatmap_std}
\end{subfigure}

\caption{Average (\subref{fig:ss_influence_mcdropout}) and standard deviation (\subref{fig:ss_influence_ensemble})
of the Ensemble's supervised accuracy $\overline{acc}$ with $\epsilon=0.1$ on OOD data, computed for the Traffic subject inside a 5x5 neighbourhood of each configuration}
\label{fig:heatmaps}
\end{figure}
The number of training epochs and the number of samples/atomic models can be visualized as a 200 (number of epochs) by 100 (number of samples) or 50 (number of atomic models) grid. 
To assess the sensitivity of supervisors, we calculate the standard deviation (\textit{std}) of $\overline{ACC}$ within a 5x5 filter applied to this grid. In this way, we account for the variability of the supervised accuracy in a 5x5 neighbourhood of each hyperparameter configuration.
The higher the std, the higher the sensitivity to small hyperparameter changes in the neighbourhood.
We also calculate the average $\overline{ACC}$ in such 5x5 neighbourhood.
The result is a pair of heatmaps, an example of which is shown in \autoref{fig:heatmaps}. 
This figure suggests that hyperparameter sensitivity negatively correlates with $\overline{ACC}$: 
Hyperparameters leading to low accuracy (bright colors in \autoref{fig:heatmap_mean}) typically show high sensitivity to hyperparameter changes (dark colors in \autoref{fig:heatmap_std}).
We do indeed observe this negative correlation for all case studies.
The values of point-biserial correlation, shown in the \emph{S-C} (Sensitivity-Correlation) columns of \autoref{tab:big_table}, are strongly negative in most cases, with $p$-value $< 0.05$ in 358 out of 390 cases.
In low accuracy cases, small changes to number of samples and number of epochs have a high effect on accuracy.
\begin{tcolorbox}
\textbf{Summary (RQ4)}: \textit{For a given supervisor and model, for what concerns the number of training epochs and samples used for quantification, the higher the supervised accuracy, the lower the model's sensitivity to small hyperparameter changes.} 
\end{tcolorbox}

\subsection{Lessons Learned}
Based on our answers to the RQs, we distilled the following primary lessons learned, possibly useful for a practical usage of 
uncertainty-quantifiers based uncertainty monitoring for DLS robustness:

\begin{description}
    \item [Anything is better than nothing:]
    While the selection of the ideal supervisor is problem dependent, 
    all the supervisors we tested showed some capability to increase the accuracy of the  DNN when supervised.
    Thus including any uncertainty monitoring supervisor in a DLS increases its fail-safety.
    \item [Ensembles are powerful:]
    Not only did Ensembles show the best average rank on the $S_1$ score (ignoring the pre-trained Imagenet studies), in many cases even the unsupervised accuracy increased.
    Furthermore, the relatively low number of 20 atomic models was sufficient to achieve good results in our study.
    Thus, provided sufficient system resources, we suggest software architects to use Ensembles instead of Point Predictors. On the other hand, the latter may represent a good compromise solution when computational resources are severely constrained.
    \item [Number of samples affects choice of quantifier:]
    For Ensembles and MC-Dropout, VR was the best quantifier on average, but it requires a large number of samples to allow for precise threshold selection. 
    Thus, if computational resources allow the calculation of a high number of samples, our experiments suggest to use VR as quantifier. Otherwise, MS showed good performance, despite its simplicity.
    \item [In-production supervisor assessment is needed:]
    Since there is no uncertainty quantifier which performs best in all cases,
    we want to emphasize the importance of in-production assessment of the supervisors performance on the actual system to be supervised, by comparing different supervisors for optimal selection. The metrics that we propose in Section~\ref{sec:assessment} are specifically designed for such assessment.
\end{description}

\subsection{Threats to Validity}

\textbf{External Validity}: While we considered only four subjects, we diversified them as much as possible. In particular, besides the benchmark subjects often used in DNN testing (Mnist, Cifar10 and Imagenet), we included an additional subject, Traffic,  which consists of unbalanced data, partially of low quality. It implements a functionality (traffic sign recognition) commonly integrated in autonomous vehicles. 

\textbf{Internal Validity}: 
The selection of hyperparameters for DNN training
might be critical and the selected values may not be representative of other contexts.
To address this threat, we refrained from selecting any  hyperparameter for the case study models ourselves, wherever possible, and instead relied on  architectures available from the literature.
For what concerns the internal hyperparameters of the supervisors, we evaluated the sensitivity of the results to their choice in a dedicated research question (RQ4).

Another threat to the internal validity of our study is that the OOD inputs used in our experiments might not be representative of the uncertainties that may be observed in practice. 
Indeed, this is unavoidable and intrinsic to the problem of DNN supervision, as unexpected conditions occurring in practice cannot be by definition simulated ahead of time.

\section{Related Work}
\label{sec:related}

\textbf{Empirical Studies of Uncertainty-Aware Deep Neural Networks:}
Oliveira \etal \cite{Oliveira2016} compared, amongst others, MC-Dropout based uncertainty and a variational approximation of BNN. 
In their experiments, the performance of the two were comparable, but MC-Dropout was much faster. 
They did not consider Ensemble models.
Similarly, but more extensively, Ovadia \etal \cite{Ovadia2019} compared various uncertainty aware DNNs against each other, including MC-Dropout,
Deep Ensembles and a variational approximation of BNN. Consistently with our results,  Ensembles performed the best in their experiments.
As opposed to our work, both of these studies consider fewer  subjects, do not investigate the impact of different quantifiers,
and do not have the constraint of transparently introducing uncertainty estimators into DNNs without altering their inner architecture, as possible instead with variational BNN approximations.
Zhang \etal \cite{Zhang2020} compare MC-Dropout against PCS to detect misclassifications caused by adversarial examples,
i.e., examples deliberately modified to trick the DNN into making prediction errors. 
Their results show, similarly to ours, that there is no strict dominance between these two approaches of network supervision. Ours is the first large scale study where uncertainty estimators are injected transparently into existing DNNs (thanks to \tool). We are also the first to introduce practical metrics for in-production assessment of supervisors and to distill a list of lessons learned that can be used as guidelines for practical usage of supervisors.

\textbf{Other types of DNN supervisors:}
While our focus is on uncertainty measures based on the variability of the output for a given input,
there have been various proposals of supervisors that are not directly based on the output distribution of the supervised network.
Berend \etal\cite{Berend2020} compared various such techniques, typically based on neuron activations, which re\-co\-gnize activation patterns that were not sufficiently represented in the training data. While such an approach may be powerful against epistemic uncertainty, it can not help against aleatoric uncertainty.
Stocco \etal\cite{Stocco2020} and Henriksson \etal\cite{Henriksson2019} proposed the use of autoencoders, i.e., anomaly detectors trained on the DLS training set as DNN supervisors. This approach does not consider the inner state of the DNN or its predictions (i.e., it is black-box). On the contrary, the supervisors supported by \tool take advantage of the predictions of the DNN being supervised.

\section{Conclusion}
\label{sec:conclusion}

Despite their fundamental role to support fail-safe execution, uncertainty estimators are rarely used by developers when integrating DNNs into complex DLS. 
This might be due to the high complexity of some of the approaches and the lack of a tool to facilitate the use of such techniques.
This paper closes such gap through the following contributions: 
(1) We compared the most widely used uncertainty-aware DNN types, providing an easy start into the relevant literature;
(2) We released our tool \tool, which allows to build and evaluate uncertainty-aware DNNs obtained transparently from unchanged, traditional DNNs;
(3) We reported our empirical evaluation, summarized into practical guidelines on how to set up an uncertainty monitoring DNN supervisor for a production system. 

The non-optimality of any approach under all conditions may allow a complementary use of multiple supervisors.
To this extent, we plan to investigate the combination of different supervisors as future work, 
hopefully leading to an overall more stable and universally applicable supervision.

\bibliographystyle{IEEEtran}
\bibliography{main}

\begin{thebibliography}{10}
\providecommand{\url}[1]{#1}
\csname url@samestyle\endcsname
\providecommand{\newblock}{\relax}
\providecommand{\bibinfo}[2]{#2}
\providecommand{\BIBentrySTDinterwordspacing}{\spaceskip=0pt\relax}
\providecommand{\BIBentryALTinterwordstretchfactor}{4}
\providecommand{\BIBentryALTinterwordspacing}{\spaceskip=\fontdimen2\font plus
\BIBentryALTinterwordstretchfactor\fontdimen3\font minus
  \fontdimen4\font\relax}
\providecommand{\BIBforeignlanguage}[2]{{%
\expandafter\ifx\csname l@#1\endcsname\relax
\typeout{** WARNING: IEEEtran.bst: No hyphenation pattern has been}%
\typeout{** loaded for the language `#1'. Using the pattern for}%
\typeout{** the default language instead.}%
\else
\language=\csname l@#1\endcsname
\fi
#2}}
\providecommand{\BIBdecl}{\relax}
\BIBdecl

\bibitem{Templeton2020Forbes}
\BIBentryALTinterwordspacing
B.~Templeton. (2020) Tesla in taiwan crashes directly into overturned truck,
  ignores pedestrian, with autopilot on. [Online]. Available:
  https://www.forbes.com/sites/bradtempleton/2020/06/02/tesla-in-taiwan-crashes-directly-into-overturned-truck-ignores-pedestrian-with-autopilot-on
 \BIBentrySTDinterwordspacing

\bibitem{Vincent2018GoogleFotos}
\BIBentryALTinterwordspacing
J.~Vincent. (2018) Google ‘fixed’ its racist algorithm by removing gorillas
  from its image-labeling tech. [Online]. Available:
https://www.theverge.com/2018/1/12/16882408/google-racist-gorillas-photo-recognition-algorithm-ai
\BIBentrySTDinterwordspacing

\bibitem{Kendall2017a}
A.~Kendall and Y.~Gal, ``What uncertainties do we need in bayesian deep
  learning for computer vision?'' in \emph{Advances in Neural Information
  Processing Systems 30}, I.~Guyon, U.~V. Luxburg, S.~Bengio, H.~Wallach,
  R.~Fergus, S.~Vishwanathan, and R.~Garnett, Eds.\hskip 1em plus 0.5em minus
  0.4em\relax Curran Associates, Inc., 2017, pp. 5574--5584.

\bibitem{Riccio2020}
V.~Riccio, G.~Jahangiroba, A.~Stocco, N.~Humbatova, M.~Weiss, and P.~Tonella,
  ``Testing machine learning based systems: a systematic mapping,''
  \emph{Empirical Software Engineering}, 2020.

\bibitem{Jospin2020}
L.~V. Jospin, W.~Buntine, F.~Boussaid, H.~Laga, and M.~Bennamoun, ``Hands-on
  bayesian neural networks -- a tutorial for deep learning users,'' 2020.

\bibitem{Ovadia2019}
Y.~Ovadia, E.~Fertig, J.~Ren, Z.~Nado, D.~Sculley, S.~Nowozin, J.~Dillon,
  B.~Lakshminarayanan, and J.~Snoek, ``Can you trust your models uncertainty?
  evaluating predictive uncertainty under dataset shift,'' \emph{Advances in
  Neural Information Processing Systems}, pp. 13\,991--14\,002, 2019.

\bibitem{Neal1992}
R.~M. Neal, ``Bayesian training of backpropagation networks by the hybrid monte
  carlo method,'' Citeseer, Tech. Rep., 1992.

\bibitem{MacKay1992}
D.~J. MacKay, ``A practical bayesian framework for backpropagation networks,''
  \emph{Neural computation}, vol.~4, no.~3, pp. 448--472, 1992.

\bibitem{Ilg2018}
E.~Ilg, O.~Cicek, S.~Galesso, A.~Klein, O.~Makansi, F.~Hutter, and T.~Brox,
  ``Uncertainty estimates and multi-hypotheses networks for optical flow,'' in
  \emph{Proceedings of the European Conference on Computer Vision (ECCV)},
  2018, pp. 652--667.

\bibitem{Gal2016}
Y.~Gal and Z.~Ghahramani, ``Dropout as a bayesian approximation: Representing
  model uncertainty in deep learning,'' in \emph{Proceedings of the 33rd
  International Conference on International Conference on Machine Learning -
  Volume 48}, ser. ICML'16.\hskip 1em plus 0.5em minus 0.4em\relax JMLR.org,
  2016, pp. 1050--1059.

\bibitem{Srivastava2014}
\BIBentryALTinterwordspacing
N.~Srivastava, G.~Hinton, A.~Krizhevsky, I.~Sutskever, and R.~Salakhutdinov,
  ``Dropout: A simple way to prevent neural networks from overfitting,''
  \emph{Journal of Machine Learning Research}, vol.~15, no.~56, pp. 1929--1958,
  2014. [Online]. Available:
  \url{http://jmlr.org/papers/v15/srivastava14a.html}
\BIBentrySTDinterwordspacing

\bibitem{Osband2016}
I.~Osband, ``Risk versus uncertainty in deep learning: Bayes, bootstrap and the
  dangers of dropout,'' in \emph{Neural Information Processing Systems (NIPS)},
  2016.

\bibitem{Lakshminarayanan2017}
B.~Lakshminarayanan, A.~Pritzel, and C.~Blundell, ``Simple and scalable
  predictive uncertainty estimation using deep ensembles,'' in \emph{Advances
  in neural information processing systems}, 2017, pp. 6402--6413.

\bibitem{Pearce2020}
T.~Pearce, F.~Leibfried, and A.~Brintrup, ``Uncertainty in neural networks:
  Approximately bayesian ensembling,'' in \emph{International conference on
  artificial intelligence and statistics}.\hskip 1em plus 0.5em minus
  0.4em\relax PMLR, 2020, pp. 234--244.

\bibitem{Berend2020}
D.~Berend, X.~Xie, L.~Ma, L.~Zhou, Y.~Liu, C.~Xu, and J.~Zhao, ``Cats are not
  fish: Deep learning testing calls for out-of-distribution awareness,'' in
  \emph{The 35th IEEE/ACM International Conference on Automated Software
  Engineering}.\hskip 1em plus 0.5em minus 0.4em\relax New York, NY, USA:
  Association for Computing Machinery, 2020.

\bibitem{Zhang2020}
X.~Zhang, X.~Xie, L.~Ma, X.~Du, Q.~Hu, Y.~Liu, J.~Zhao, and M.~Sun, ``Towards
  characterizing adversarial defects of deep learning software from the lens of
  uncertainty,'' in \emph{Proceedings of 42nd International Conference on
  Software Engineering}.\hskip 1em plus 0.5em minus 0.4em\relax ACM, 2020.

\bibitem{Nix1994}
D.~A. Nix and A.~S. Weigend, ``Estimating the mean and variance of the target
  probability distribution,'' in \emph{Proceedings of 1994 ieee international
  conference on neural networks (ICNN'94)}, vol.~1.\hskip 1em plus 0.5em minus
  0.4em\relax IEEE, 1994, pp. 55--60.

\bibitem{Gal2016a}
Y.~Gal, ``Uncertainty in deep learning,'' Ph.D. dissertation, University of
  Cambridge, 2016.

\bibitem{Michelmore2018}
R.~Michelmore, M.~Kwiatkowska, and Y.~Gal, ``Evaluating uncertainty
  quantification in end-to-end autonomous driving control,'' \emph{CoRR}, 2018.

\bibitem{TfPytorchCompare}
\BIBentryALTinterwordspacing
GithubCompare, \emph{Github statistics comparison between tensorflow and
  pytorch}, Accessed September 25, 2020. [Online]. Available:
  \url{https://www.githubcompare.com/tensorflow/tensorflow+pytorch/pytorch}
\BIBentrySTDinterwordspacing

\bibitem{Weiss2020Wizard}
M.~Weiss and P.~Tonella, ``Uncertainty-wizard: Fast and user-friendly neural
  network uncertainty quantification,'' \emph{arXiv preprint arXiv:2101.00982},
  2020.

\bibitem{Stocco2020}
A.~Stocco, M.~Weiss, M.~Calzana, and P.~Tonella, ``Misbehaviour prediction for
  autonomous driving systems,'' in \emph{Proceedings of 42nd International
  Conference on Software Engineering}.\hskip 1em plus 0.5em minus 0.4em\relax
  ACM, 2020, p. 12 pages.

\bibitem{Henriksson2019}
J.~Henriksson, C.~Berger, M.~Borg, L.~Tornberg, C.~Englund, S.~R.
  Sathyamoorthy, and S.~Ursing, ``Towards structured evaluation of deep neural
  network supervisors,'' in \emph{2019 {IEEE} International Conference On
  Artificial Intelligence Testing ({AITest})}.\hskip 1em plus 0.5em minus
  0.4em\relax {IEEE}, apr 2019.

\bibitem{Davis2006}
J.~Davis and M.~Goadrich, ``The relationship between precision-recall and roc
  curves,'' in \emph{Proceedings of the 23rd international conference on
  Machine learning - ICML06}.\hskip 1em plus 0.5em minus 0.4em\relax {ACM}
  Press, 2006.

\bibitem{Saito2015}
T.~Saito and M.~Rehmsmeier, ``The precision-recall plot is more informative
  than the {ROC} plot when evaluating binary classifiers on imbalanced
  datasets,'' \emph{{PLOS} {ONE}}, vol.~10, no.~3, p. e0118432, mar 2015.

\bibitem{Rijsbergen1979}
\BIBentryALTinterwordspacing
C.~van Rijsbergen, \emph{Information Retrieval, 2nd Edition, Chapter 7}.\hskip
  1em plus 0.5em minus 0.4em\relax Butterworths, 1979. [Online]. Available:
  \url{http://www.dcs.gla.ac.uk/Keith/Preface.html}
\BIBentrySTDinterwordspacing

\bibitem{LeCun1998}
Y.~LeCun, L.~Bottou, Y.~Bengio, and P.~Haffner, ``Gradient-based learning
  applied to document recognition,'' \emph{Proceedings of the IEEE}, vol.~86,
  no.~11, pp. 2278--2324, 1998.

\bibitem{KerasMnistModel}
\BIBentryALTinterwordspacing
Keras-Team. (2018) Code: Convolutional neural network example. [Online].
  Available:
  \url{https://github.com/keras-team/keras/blob/master/examples/mnist_cnn.py}
\BIBentrySTDinterwordspacing

\bibitem{Krizhevsky2009}
A.~Krizhevsky, G.~Hinton \emph{et~al.}, ``Learning multiple layers of features
  from tiny images,'' 2009.

\bibitem{BrownleeCifarModel}
\BIBentryALTinterwordspacing
J.~Brownlee. (2019) How to develop a cnn from scratch for cifar-10 photo
  classification. [Online]. Available:
  https://machinelearningmastery.com/how-to-develop-a-cnn-from-scratch-for-cifar-10-photo-classification/
\BIBentrySTDinterwordspacing

\bibitem{Serna2018}
C.~G. Serna and Y.~Ruichek, ``Classification of traffic signs: The european
  dataset,'' \emph{IEEE Access}, 2018.

\bibitem{Segvic2010}
S.~{\v{S}}egvic, K.~Brki{\'c}, Z.~Kalafati{\'c}, V.~Stanisavljevi{\'c},
  M.~{\v{S}}evrovi{\'c}, D.~Budimir, and I.~Dadi{\'c}, ``A computer vision
  assisted geoinformation inventory for traffic infrastructure,'' in \emph{13th
  International IEEE Conference on Intelligent Transportation Systems}.\hskip
  1em plus 0.5em minus 0.4em\relax IEEE, 2010, pp. 66--73.

\bibitem{Bonaci2011}
I.~Bonaci, I.~Kusalic, I.~Kovacek, Z.~Kalafatic, and S.~Segvic, ``Addressing
  false alarms and localization inaccuracy in traffic sign detection and
  recognition,'' in \emph{16th computer vision winter workshop}.\hskip 1em plus
  0.5em minus 0.4em\relax Citeseer, 2011, pp. 1--8.

\bibitem{Mathias2013}
M.~Mathias, R.~Timofte, R.~Benenson, and L.~Van~Gool, ``Traffic sign
  recognition—how far are we from the solution?'' in \emph{The 2013
  international joint conference on Neural networks (IJCNN)}.\hskip 1em plus
  0.5em minus 0.4em\relax IEEE, 2013, pp. 1--8.

\bibitem{Timofte2014}
R.~Timofte, K.~Zimmermann, and L.~Van~Gool, ``Multi-view traffic sign
  detection, recognition, and 3d localisation,'' \emph{Machine vision and
  applications}, vol.~25, no.~3, pp. 633--647, 2014.

\bibitem{Belaroussi2010}
R.~Belaroussi, P.~Foucher, J.-P. Tarel, B.~Soheilian, P.~Charbonnier, and
  N.~Paparoditis, ``Road sign detection in images: A case study,'' in
  \emph{2010 20th International Conference on Pattern Recognition}.\hskip 1em
  plus 0.5em minus 0.4em\relax IEEE, 2010, pp. 484--488.

\bibitem{Stallkamp2012}
J.~Stallkamp, M.~Schlipsing, J.~Salmen, and C.~Igel, ``Man vs. computer:
  Benchmarking machine learning algorithms for traffic sign recognition,''
  \emph{Neural networks}, vol.~32, pp. 323--332, 2012.

\bibitem{Grigorescu2003}
C.~Grigorescu and N.~Petkov, ``Distance sets for shape filters and shape
  recognition,'' \emph{IEEE transactions on image processing}, vol.~12, no.~10,
  pp. 1274--1286, 2003.

\bibitem{Larsson2011}
F.~Larsson and M.~Felsberg, ``Using fourier descriptors and spatial models for
  traffic sign recognition,'' in \emph{Scandinavian conference on image
  analysis}.\hskip 1em plus 0.5em minus 0.4em\relax Springer, 2011, pp.
  238--249.

\bibitem{Deng2009}
J.~Deng, W.~Dong, R.~Socher, L.-J. Li, K.~Li, and L.~Fei-Fei, ``Imagenet: A
  large-scale hierarchical image database,'' in \emph{2009 IEEE conference on
  computer vision and pattern recognition}.\hskip 1em plus 0.5em minus
  0.4em\relax Ieee, 2009, pp. 248--255.

\bibitem{Tan2019}
M.~Tan and Q.~V. Le, ``Efficientnet: Rethinking model scaling for convolutional
  neural networks,'' \emph{arXiv preprint arXiv:1905.11946}, 2019.

\bibitem{Mu2019}
N.~Mu and J.~Gilmer, ``Mnist-c: A robustness benchmark for computer vision,''
  \emph{arXiv}, 2019.

\bibitem{Hendrycks2018}
D.~Hendrycks and T.~Dietterich, ``Benchmarking neural network robustness to
  common corruptions and perturbations,'' \emph{International Conference on
  Learning Representations}, 2018.

\bibitem{Oliveira2016}
R.~Oliveira, P.~Tabacof, and E.~Valle, ``Known unknowns: Uncertainty quality in
  bayesian neural networks,'' \emph{Workshop on Bayesian Deep Learning, NIPS
  2016, Barcelona, Spain}, 2016.

\end{thebibliography}

\end{document}